\documentclass[twocolumn,pra,twocolumn,superscriptaddress]{revtex4}
\usepackage{color}
\usepackage{amsmath,amsthm,mathrsfs,dsfont}
\usepackage{graphicx}
\usepackage{tikz}
\usepackage{xr}
\usepackage{xc}
\usepackage{float}
\usepackage{bbm}
\usepackage{epsfig}
\usepackage{verbatim}
\usepackage{array}
\usepackage{setspace}
\usepackage{bbding}
\usepackage{amssymb}
\usepackage{pifont}
\usepackage{braket}
\usepackage{tikz}
\usepackage{qcircuit}

\usepackage{newtxtext,newtxmath}

\externalcitedocument[suppl-]{../suppl/low-depth0312}

\usepackage[unicode=true,
 bookmarks=true,bookmarksnumbered=false,bookmarksopen=false,
 breaklinks=false,pdfborder={0 0 1},backref=false,colorlinks=true]
 {hyperref}
\hypersetup{
 linkcolor=magenta, urlcolor=blue, citecolor=blue, pdfstartview={FitH}, hyperfootnotes=false, unicode=true}

\usepackage{color}

\makeatletter
\@ifundefined{textcolor}{}
{%
 \definecolor{BLACK}{gray}{0}
 \definecolor{WHITE}{gray}{1}
 \definecolor{RED}{rgb}{1,0,0}
 \definecolor{GREEN}{rgb}{0,1,0}
 \definecolor{BLUE}{rgb}{0,0,1}
 \definecolor{CYAN}{cmyk}{1,0,0,0}
 \definecolor{MAGENTA}{cmyk}{0,1,0,0}
 \definecolor{YELLOW}{cmyk}{0,0,1,0}
}

\usepackage{amsfonts}\usepackage{tabularx}\usepackage{dcolumn}\usepackage{bm}\usepackage{graphicx}\usepackage{epstopdf}

\hypersetup{urlcolor=blue}

\makeatother

\newcolumntype{C}[1]{>{\centering\arraybackslash$}p{#1}<{$}}

\usepackage{algorithm}
\usepackage{algorithmic}

\begin{document}

\widetext
\title{Robust M\o lmer-S\o rensen Gate Against Symmetric and Asymmetric Errors}
\author{Wenhao Zhang}
\thanks{These two authors contributed equally}
\affiliation{Center on Frontiers of Computing Studies, School of Computer Science, Peking University, Beijing 100871, China}
 \author{Gaoxiang Tang}
 \thanks{These two authors contributed equally}
\affiliation{Center for Quantum Information, IIIS, Tsinghua University, Beijing 100084, People's Republic of China}
 \author{Kecheng Liu}
 \affiliation{Center on Frontiers of Computing Studies, School of Computer Science, Peking University, Beijing 100871, China}
 \author{Xiao Yuan}
 \affiliation{Center on Frontiers of Computing Studies, School of Computer Science, Peking University, Beijing 100871, China}
 \author{Yangchao Shen}
 \email{shenyangchao@gmail.com}
 \affiliation{Center on Frontiers of Computing Studies, School of Computer Science, Peking University, Beijing 100871, China}
 \author{Yukai Wu}
 \email{wyukai@mail.tsinghua.edu.cn}
\affiliation{Center for Quantum Information, IIIS, Tsinghua University, Beijing 100084, People's Republic of China}
\affiliation{Hefei National Laboratory, Hefei 230088, PR China}
\author{Xiao-Ming Zhang}
\email{phyxmz@gmail.com}
\affiliation{Center on Frontiers of Computing Studies, School of Computer Science, Peking University, Beijing 100871, China}
\affiliation{Key Laboratory of Atomic and Subatomic Structure and Quantum Control (Ministry of Education), South China Normal University, Guangzhou 510006, China}
\affiliation{Guangdong Provincial Key Laboratory of Quantum Engineering and Quantum Materials, Guangdong-Hong Kong Joint Laboratory of Quantum Matter, South China Normal University, Guangzhou 510006, China}

\begin{abstract}
To achieve the entangling gate fidelity above the quantum error correction threshold, it is critical to suppress errors due to experimental imperfection. We consider the M\o lmer-S\o rensen gates in trapped-ion systems, and develop a general approach to suppress a family of noise sources that appeared as either symmetric or asymmetric errors. Using the time-average displacement minimization technique, both symmetric error and displacement-dependent part of the asymmetric errors are eliminated. Then, by analyzing the tangent space of displacement-independent errors, we obtain the analytic form of the generators of the correction operator to the remaining error terms. We then develop a compensation pulse to fully suppress the remaining displacement-independent errors. The effectiveness of our scheme is further verified by numerical analysis, through which we observe a significant reduction of entangling gate infidelity. Our findings enhance gate fidelity and robustness to noise for ion trap quantum computing.
\end{abstract}
\maketitle
\section{Introduction}
Trapped-ion systems are among the most promising platforms for quantum computing, offering several advantages such as long coherence times of a few seconds \cite{kirchhoff2024correction}, high-accuracy qubit initialization, addressing, and all-to-all coupling topology. The quantum gate fidelity is also among the highest, a single-qubit gate fidelity above $99.9999\%$ \cite{PhysRevLett.113.220501, smith2024singlequbitgateserrors107}, a two-qubit gate fidelity above $99.9\%$ \cite{PhysRevLett.117.060504,PhysRevLett.117.060505,PhysRevLett.127.130505}, and a state-preparation-and-measurement fidelity above $99.99\%$ \cite{PhysRevLett.129.130501} have been reported.

Recent demonstrations of reliable quantum computing tasks on trapped-ion quantum computers with up to 32 ions \cite{Hong_2024} indicate rapid progress in this field. Techniques such as QCCD \cite{kielpinski_architecture_2002}, ion-photon networks \cite{RevModPhys.82.1209}, and 2D ion crystals \cite{wu2024progress} have the potential to further increase the number of ions. These characteristics make trapped-ion systems strong candidates for realizing fault-tolerant quantum computing and for near-term noisy intermediate-scale quantum devices with enhanced performance.

In trapped-ion quantum computers, ions are confined in one-dimensional or two-dimensional lattices formed by the pseudopotential of Paul traps. Each physical qubit is encoded in the energy levels of an individual ion, such as hyperfine levels or optical levels \cite{10.1063/1.5088164}. The small vibrations of the ions around their equilibrium positions are correlated by the Coulomb interaction between them, hence can be described in terms of various collective motional modes and further quantized as phonons. Single-qubit gates are implemented via resonant driving or stimulated Raman transitions \cite{campbell2010ultrafast}. Entanglement between qubits is mediated by phonon modes. The primary two-qubit gate schemes in use are the $\sigma_z$-type Light-Shift gate \cite{milburn2000ion,leibfried2003experimental} and the $\sigma_\phi$-type M\o lmer-S\o rensen (MS) gate \cite{molmer1999multiparticle}. In recent years, MS gates have been extensively used in trapped-ion systems \cite{shapira2018robust,leung2018robust,figgatt2019parallel, lu2019global,PRXQuantum.1.020316,duan2022robust,hou2024individually}, and their fidelity still remains the bottleneck for achieving high-fidelity quantum computing on this platform.

The original MS gate protocol involves tuning the laser detuning near a specific phonon mode to facilitate entanglement through that mode \cite{molmer1999multiparticle}. Subsequent developments have considered multiple phonon modes and required that they decouple from the qubit states at the end of the gate operation, which necessitates modulation of the laser fields used to implement the gate through different modulation techniques~\cite{zhu2006arbitrary,roos2008ion,green2015phase,milne2020phase,leung2018robust}. In practical implementations, the MS gate may be vulnerable to various noise sources \cite{wu2018noise,PhysRevLett.119.220505,PhysRevA.107.062406,martinez2022analytical}. For example, the drifts of trap frequency can result in symmetric errors that can be effectively modeled as the trajectory drift in phase spaces. Fortunately, the symmetry errors can be compensated, up to the first order, by a robust waveform design~\cite{leung2018robust}, which is referred to as the symmetry error-robust waveform design in the reminder of this work.

Apart from those symmetric errors, other noise sources may emerge as asymmetric errors. For example, unwanted external magnetic fields will cause the fluctuation of the frequency of the qubits encoded in Zeeman levels and hence the qubit energy splitting. In experiments, the unwanted magnetic fields have been reported as the second largest error source for \( ^{40}\mathrm{Ca}^+ \) qubits, limiting their $T_2$ time to the level of \(10\)–\(100 \, \mathrm{ms} \) \cite{pogorelov2021compact,Postler_2022}. For \( ^{88}\mathrm{Sr}^+ \) qubits \cite{manovitz2022trapped}, the asymmetric $\sigma_z$-type error is also the dominant error source. Even in other qubits encoded in clock transitions that are insensitive to magnetic field fluctuations, such as \( ^{171}\mathrm{Yb}^+ \) hyperfine qubits \cite{wang2020high}, asymmetric errors will still emerge from laser frequency miscalibrations. The impact of the so-called \textit{center-line miscalibration} has been examined systematically in previous research \cite{martinez2022analytical}. Based on Magnus expansion analysis and experimental verification, Ref.~\cite{martinez2022analytical} showed that the asymmetric error is non-negligible. Unfortunately, the asymmetric errors mentioned above cannot be directly suppressed by existing error mitigation approaches. New techniques are thus required to achieve high robustness against asymmetric errors.

In this work, we propose an MS gate scheme designed to be intrinsically robust against asymmetric errors, symmetric errors, and single-qubit errors. We first show that the displacement-dependent part of the asymmetric error can be suppressed by the symmetry error-robust waveform design as well. The remaining displacement-independent part of errors is suppressed by a compensation pulse, which is designed using our Generator-Based Compensation (GBC) technique. This scheme is also robust to general noise sources beyond asymmetric errors, including symmetric errors and single-qubit errors. The robustness of our scheme is also verified with numerical simulations.

The remainder of this paper is organized as follows. In Sec.~\ref{sec:MS_gate}, we review the current scheme for implementing the MS gate using amplitude-modulated laser pulses and discuss the effects of asymmetric errors under experimental conditions. In Sec.~\ref{sec:methods}, we present our gate design scheme, providing both algebraic analysis and geometric illustrations. We also detail the quantum circuit implementation. In Sec.~\ref{sec:numerics}, we provide numerical results that showcase the advantages of our approach. Finally, in Sec.~\ref{sec:conclusion}, we conclude with discussions and further remarks on our findings.

\begin{figure*}
    \centering
    \includegraphics[width=0.85\textwidth]{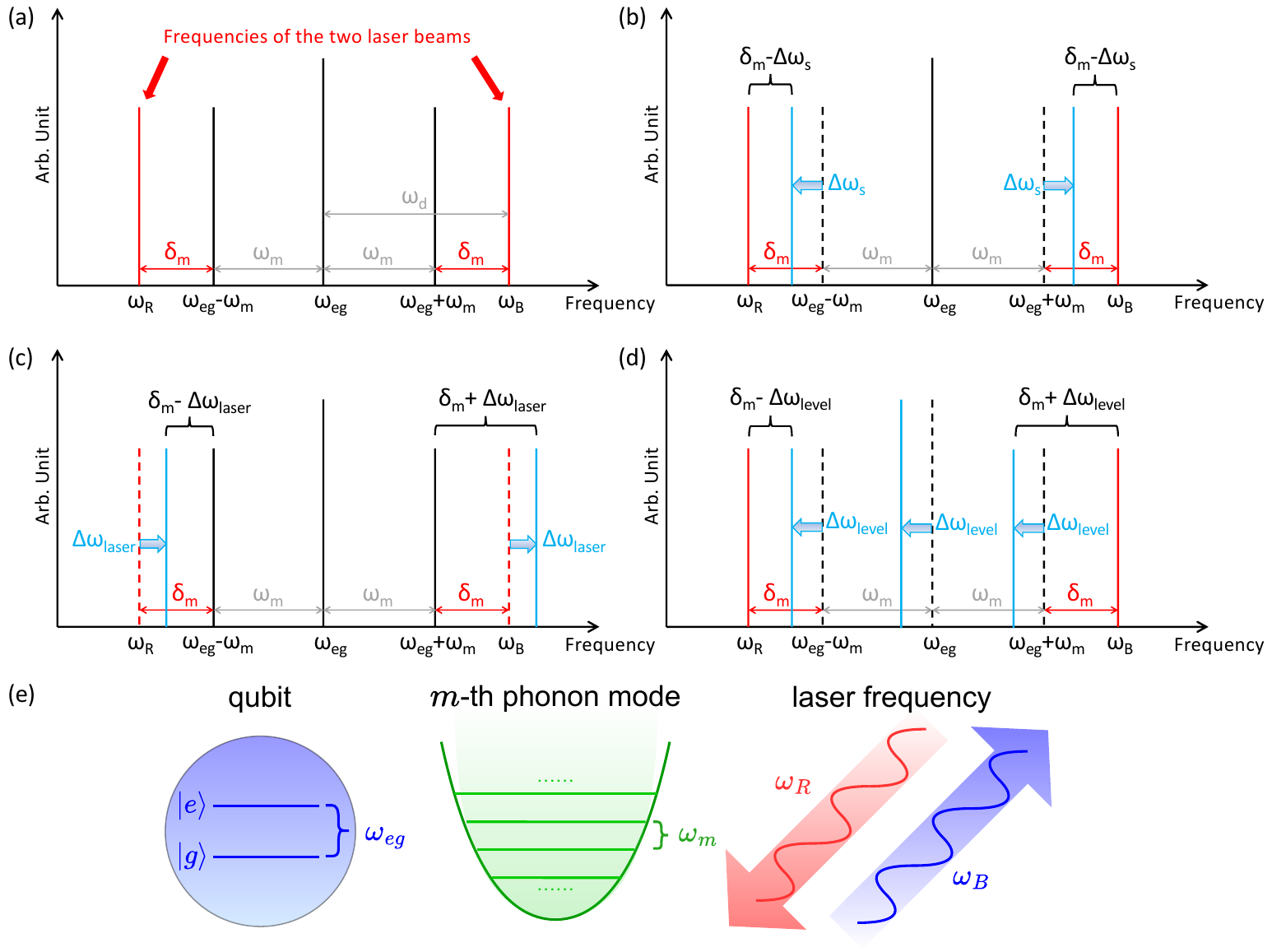}
    \caption{Drift patterns of frequencies that cause different types of coherent errors. (a) Ideal setup of laser frequencies, which could be impaired by: (b) symmetric error; (c) asymmetric error caused by laser frequency shift; (d) asymmetric error caused by qubit frequency shift. In sub-figures (a)-(d), $\omega_{\rm eg}$ denotes the internal frequency of the ion, $\omega_{m}$ denotes the phonon frequency of the $m$-th vibrational mode, $\omega_R=\omega_{\rm eg}-\omega_d$ and $\omega_B=\omega_{\rm eg}+\omega_d$ represent the red-detuned and blue-detuned laser frequencies, which are supposed to symmetrically deviate from $\omega_{\rm eg} - \omega_{m}$ and $\omega_{\rm eg} + \omega_{m}$ by detuning $\delta_{m}$, respectively, as illustrated in (a). The $\Delta\omega_s, \Delta\omega_{\rm laser}, \Delta\omega_{\rm level}$ in (b), (c) and (d) represent the corresponding frequency shift, which in the symmetric case (b) changes the ideal detuning from $\delta_m$ to $\delta_m - \Delta\omega_s$ for both red and blue detuned lasers, while in the asymmetric cases (c) and (d) changes the detuning to $\delta_m - \Delta\omega_{\rm laser(level)}$ and $\mu + \Delta\omega_{\rm laser(level)}$ for red and blue detuned lasers. (e) Sketch of the definitions of qubit frequency $\omega_{\text{eg}}$, phonon mode $\omega_m$, and laser frequencies $\omega_{R,B}$.}
    \label{Fig_errors}
\end{figure*}

\section{Asymmetric Errors in MS Gate Dynamics}
\label{sec:MS_gate}

\subsection{Ideal Hamiltonian}

Consider $N$ ions in a linear Paul trap along the z-axis. For typical experimental parameters, the micromotion is small and can be neglected. The free Hamiltonian of the system is

\begin{equation}
\label{eq:free_ham}
    \hat{H}_0 =\sum_{j=1}^{N}\frac{\omega_{eg}}{2}\sigma^{j}_z+\sum_{m=1}^N \omega_m (a_m^\dagger a_m+\frac{1}{2}),
\end{equation}

Where $\omega_{eg}$ is the energy gap between two internal states used as computation basis, and $\omega_m$ is the frequency of the $m$-th vibrational mode of the ion crystal.

For hyperfine qubits, MS gate is implemented through stimulated Raman transitions \cite{campbell2010ultrafast}. For optical qubits \cite{pogorelov2021compact}, MS gate is realized by applying a bichromatic laser on two ions. We will focus on optical qubits in the main text for simplicity. The noise source for the stimulated Raman transition scheme, which could be modeled in an analogous way, is discussed in Appendix.~\ref{app:SR_error}.


For optical qubits, the Hamiltonian describing the interaction with the bichromatic laser field is given by:

\begin{equation}
    \hat{H}^\prime = \sum_{\mu=\text{R,B}}\sum_{j=j_1,j_2}\frac{\Omega(t)}{2}\sigma^{j}_x \cos(\mathbf{k}\cdot\mathbf{x}_j-\omega_\mu t+\phi_\mu),
\end{equation}
where we have assumed that the bichromatic laser beams shined on both ions are identical, with their time-dependent Rabi frequency denoted by $\Omega(t)$ for both red- and blue-detuned components. These two components have different frequencies, denoted by $\omega_R$, $\omega_B$, and initial phases denoted by $\phi_R$, $\phi_B$, for red-detuned and blue-detuned, respectively. $\mathbf{k}$ is the wave vector of the bichromatic laser. Ideally, we have $\omega_R = \omega_{eg} - \omega_d$ and $\omega_B = \omega_{eg} + \omega_d$.

The Hamiltonian in the interaction picture defined by $H_0$ is
\begin{equation}
    \begin{aligned}
    \hat{H}_{\text{int}} &=e^{i\hat{H}_0t}\hat{H}^\prime e^{-i\hat{H}_0t}\\
    &=\frac{i}{2}\Omega(t)\sum_{j, m}\eta_{m}b_j^m\left(\sigma^{j}_- e^{i\phi_s}-\sigma^{j}_+ e^{-i\phi_s}\right)\\
    &\times\left(a_m^\dagger e^{i(\delta_m t - \phi_m)}+a_m e^{i(-\delta_m t+\phi_m)}\right),
\end{aligned}
\end{equation}
where $\phi_s=(\phi_R+\phi_B)/2$ is the relative phase in the spin space, $\phi_m=(\phi_R-\phi_B)/2$ is the relative phase in the motional space; the parameter $\eta_{m}=k_z \sqrt{{\hbar}/{2m\omega_m}}$ is the Lamb-Dicke parameter of the $m$-th motional mode, $b_j^m$ is the coupling coefficient between the $m$-th motional mode and the $j$-th ion, and $\delta_m=\omega_m-\omega_d$ is the detuning between the laser and the $m$-th phonon mode. Note that the Rotating Wave Approximation(RWA) and Lamb-Dicke Regime have already be applied.

By setting $\phi_R=3\pi/2$ and $\phi_B=-\pi/2$, we have $\phi_m=\pi$ and $\phi_s=\pi/2$, thus the Hamiltonian is
\begin{equation}
\label{ideal_ham}
    \hat{H}_{\text{int}} =\frac{1}{2}\Omega(t)\sum_{j, m}\eta_{m}b_j^m\sigma^{j}_{x}
    \left(a_m^\dagger e^{i\delta_m t}+a_m e^{-i\delta_m t}\right).
\end{equation}
We also point out that these phases could be tuned to other equivalent settings for the convenience of particular experiments, and different settings might be adopted in other papers. For instance, setting $\phi_m=\pi/2$ and $\phi_s=\pi/2$ recovers Eq.~(10) in \cite{wu2018noise}, and setting $\phi_m=\pi$ and $\phi_s=0$ recovers Eq.~(4) in \cite{martinez2022analytical}.

\subsection{Asymmetric error in the evolution operator}

In the MS gate implementation above, coherent frequency errors can be categorized into two types, and presented as two different terms in the Hamiltonian. The first type is symmetric errors, as illustrated in Fig.~\ref{Fig_errors}(b). Typically introduced by motional mode frequency drifts, this kind of error results in a $\tilde{\delta}_m=\delta_m-\Delta\omega_s$ in Eq.~\eqref{ideal_ham}, and can be suppressed by the symmetry error-robust waveform design\cite{leung2018robust}. The second type is asymmetric errors. As illustrated in Fig.~\ref{Fig_errors}(c)(d), it may emerge from two distinct sources: qubit level drifts and miscalibration of lasers.

In the presence of qubit energy level drifts $\Delta\omega_{\text{level}}$, the energy gap between qubit levels can be expressed as \(\omega_{eg}^\prime = \omega_{eg} + \Delta\omega_{\text{level}}\). This internal frequency drift may arises from the sensitivity of qubit energy to external magnetic fields, especially in the case of \( ^{40}\mathrm{Ca}^+ \) and  \( ^{88}\mathrm{Sr}^+ \), where magnetic field noise has become the predominant factor contributing to qubit decoherence \cite{pogorelov2021compact,Postler_2022,manovitz2022trapped}. Additionally, the AC-Stark shift also contributes to these energy level drifts. While it can be mitigated through techniques such as applying an extra laser field~\cite{PhysRevLett.90.143602} or adjusting laser detuning~\cite{PhysRevB.90.155306}, the compensation becomes increasingly difficult when using amplitude modulation, because the laser amplitude is constantly changing during the gate implementation.

Another significant source of error is the frequency drift of lasers due to miscalibration of the bichromatic laser frequencies. This bichromatic laser is generated by modulating a reference monochromatic laser that is resonant with the qubit energy gap \(\omega_{eg}\). The frequency miscalibration of the monochromatic laser will lead to the shifts of both red and blue sidebands, expressed as:
\begin{equation}
\begin{aligned}
    \omega_R &= \omega_{eg} - \omega_d + \Delta\omega_{\text{laser}}, \\
    \omega_B &= \omega_{eg} + \omega_d + \Delta\omega_{\text{laser}},
\end{aligned}
\end{equation}
where \(\Delta\omega_{\text{laser}}\) represents the frequency drift error. In an interaction picture with respect to 
\begin{small}\begin{equation}
\label{eq:miscalib_int_pic}
    H_{0}^\prime=\sum_{j}\frac{\omega_{eg}+\Delta\omega_{\text{laser}}}{2}\sigma^{j}_z+\sum_m \omega_m(a_m^\dagger a_m+\frac{1}{2})
\end{equation}
\end{small}the Hamiltonian becomes
\begin{small}\begin{equation}
\label{eq:miscalib_ham}
    \hat{H}_{\text{int}}^\prime =\frac{\Delta\omega}{2} \sum_{j}\sigma^{j}_z+ \frac{1}{2}\Omega(t) \sum_{j, m}\eta_{m}b_j^m\sigma^{j}_x
    \left(a_m^\dagger e^{i\delta_m t}+a_m e^{-i\delta_m t}\right),
\end{equation}
\end{small}where \(\Delta\omega = \Delta\omega_{\text{level}} - \Delta\omega_{\text{laser}}\) is the total asymmetric error considering both error sources. Note that the interaction picture defined by Eq.~\eqref{eq:miscalib_int_pic} differs from Eq.~\eqref{eq:free_ham}, which will also introduce a $\sigma_z$-type error during single-qubit gate operations. We will discuss how this error can be compensated in Sec.~\ref{subsec:supcode}.

Using the Magnus expansion, we are able to obtain an infinite series expression for the time evolution operator. 
We truncate the infinite series at the second order (i.e. neglecting terms at the order of $O(\Delta\omega\eta^2)$, $O(\Delta\omega^2\eta)$ and higher), with a further discussion on the higher-order terms given in Appendix.~\ref{app:higher_mag}. After the truncation, the unitary evolution operator from time $0$ to time $t$ can be expressed as
\begin{equation}
\begin{aligned}
\label{eq:2rd_mag}
\hat{U}(t)
=& \exp \left[ i \Theta(t) \sigma_x^{j_1} \sigma_x^{j_2} -\frac{i}{2} t \Delta\omega \sum_j \sigma_z^j \right.\\
& -\frac{i}{2} \sum_{j, m}\left(\alpha_m(t) a_m^{\dagger}+\alpha_m^*(t) a_m\right) \eta_{m}b_j^m \sigma_x^j \\
& \left.-\frac{i}{4} \Delta\omega \sum_{m, j}\left(\beta_m(t) a_m^{\dagger}+\beta_m^*(t) a_m\right) \eta_{m}b_j^m \sigma_y^j \right],
\end{aligned}
\end{equation}
where
\begin{equation}
\begin{aligned}
\alpha_m(t) &= \int_0^t \Omega(t_1) e^{i \delta_m t_1} \: {\rm d}t_1, \\
\beta_m(t) &= \int_0^t \alpha_m(t_1) \: {\rm d}t_1 + i \frac{\partial \alpha_m(t)}{\partial \delta_m}, \\
\Theta(t) &= \frac{1}{2}\sum_m \eta_{m}^2b_{j_1}^m b_{j_2}^m \\
&\quad \int_0^t {\rm d}t_1 \int_0^{t_1} {\rm d} t_2\:\Omega(t_1)\Omega(t_2) \sin[\delta_m(t_1-t_2)].
\end{aligned}
\end{equation}

\begin{figure*}
    \centering
    \includegraphics[width=0.85\textwidth]
    {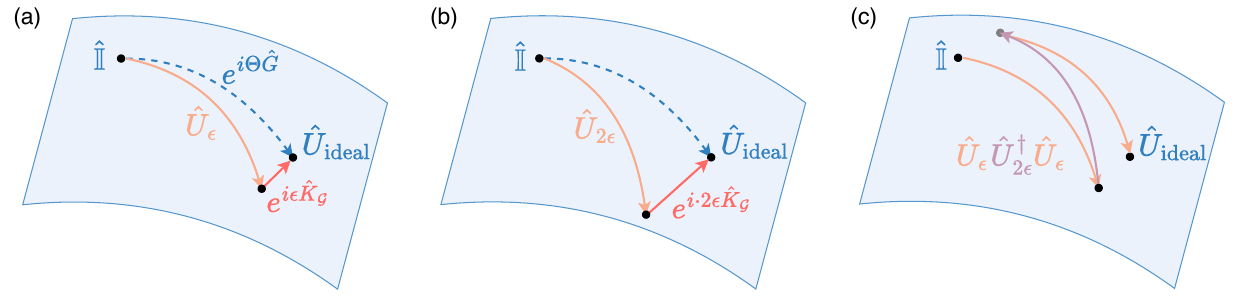}
    \caption{Illustration of the error compensation scheme based on Lie generator. (a) An error term $i\epsilon\hat{E}$ causes the ideal evolution operator to deviate a little from $\hat{U}_{\text{ideal}}$, the difference characterized by $e^{i\epsilon\hat{K}_{\mathcal{G}}}$, with $\hat{K}_{\mathcal{G}} = -i\frac{\partial}{\partial\epsilon}(\hat{U}_{\rm ideal}\hat{U}^{\dagger}_{\epsilon})|_{\epsilon = 0}$ being the generator in the corresponding tangent space. (b) When the noise level $\epsilon$ becomes $2\epsilon$ instead, the deviation would stay approximately in the same direction in the tangent space, hence writes $e^{i\cdot 2\epsilon\hat{K}_{\mathcal{G}}}$. (c) Our procedure of combining $\hat{U}_{\epsilon}$ and $\hat{U}^{\dagger}_{2\epsilon}$ to acquire error mitigated $\hat{U}_{\epsilon}^{(gbc)}$. }
    \label{Fig_tanSpace}
\end{figure*} 

Equation \eqref{eq:2rd_mag} contains four terms. The first term $\exp(i\Theta(t)\cdot \sigma_x^{j_1}\sigma_x^{j_2})$ is an entangling term between the $j_1$-th and the $j_2$-th ion, the second term $-i/2\cdot t\Delta\omega\sum_j\sigma_z^j$ is the asymmetric error term that is independent of the ion displacement, while the last two terms couple the internal states of ions with their collective motions.

\section{Robust MS gate}\label{sec:methods}
\subsection{Symmetry Error-Robust Waveform Design}

Comparing Eq.~\eqref{eq:2rd_mag} to the ideal entangling gate $\hat{U}_{\rm ideal}=\exp(i\pi/4\cdot \sigma_x^{j_1}\sigma_x^{j_2})$, we should eliminate the last three terms and keep only the first term of spin-spin entanglement. Moreover, suppose the total gate time is $\tau$, the entangling angle $\Theta(\tau)$ should be tuned to the target angle, which is set as $\Theta(\tau) = \Theta\equiv\pi/4$ in this work.

First of all, via modulating the amplitude of the external laser field, we can easily ensure that for each phonon mode $m$, $\alpha_m(\tau)=0$, thus eliminating the third term in the index of Eq.~\eqref{eq:2rd_mag} at the end of the gate. During the optimization step, the symmetry error-robust waveform design could be applied to enhance the robustness to symmetric errors. In this scheme, instead of optimizing $\alpha_m$ at $t=\tau$, we require that
    \begin{align}
        &\qquad\Omega(t) = \Omega(\tau-t)\\
        &\int_0^{\tau}\alpha_m(t)\:{\rm d}t = 0,\quad\forall m
    \end{align}

which means that the Rabi frequency of the external laser field should be symmetric during the gate implementation, and the time average of the displacement $\overline{\alpha_m}=\int_0^{\tau}\alpha_m(t)\:{\rm d}t$ vanishes. It can be proven that under these constraints, $\alpha_m(\tau)=0$ will be automatically satisfied. Detailed verification is given in Appendix.~\ref{app:amplitude}. In addition, we also show that under this scheme, $\beta_m$ is also automatically guaranteed to vanish at the end of the gate. This property suggests that we can apply the symmetry error-robust waveform design to simultaneously eliminate both displacement-dependent terms in Eq.~\eqref{eq:2rd_mag}. Hence the evolution operator is simplified as
\begin{equation}
    \hat{U}(\tau)=\exp\left(i \Theta \sigma_x^{j_1} \sigma_x^{j_2} - \frac{i}{2} \Delta\omega\tau \sum_j \sigma_z^j\right).
    \label{eq:U_tau}
\end{equation}
Therefore, our remaining task is to eliminate the displacement-independent error term $\frac{i}{2} \Delta\omega\tau \sum_j \sigma_z^j$. Although it resembles a single qubit phase shift in mathematical expression, the fact that it does not commute with the gate term hinders us from compensating it using single qubit phase gates after the MS-gate. The scheme to tackle this issue is detailed below.

\subsection{{Generator-based compensation}}
\label{subsec:dd}To rewrite the evolution operator in a more compact form, we define $\hat{G} =\sigma_x^{j_1}\sigma_x^{j_2}$ and $\hat{E} = \frac{1}{2}\sum_{j=j_1,j_2}\sigma_z^{j}$ as the target entangling operator and the asymmetric error operator, respectively. The actual evolution operator Eq.~\eqref{eq:U_tau} becomes
\begin{equation}
    \hat{U}_{\epsilon} = \exp\left(i\Theta \hat{G}-i\epsilon \hat{E}\right),
    \label{eq:U_epsilon}
\end{equation}
where $\epsilon\equiv\Delta\omega\tau$ is the error level coefficient, characterizing the strength of the asymmetric error. Our task is to recover the ideal entangling gate $\hat{U}_{\text{ideal}} = {\rm exp}(i\Theta \hat{G})$ from the noisy operator Eq.~\eqref{eq:U_epsilon}, with unknown error level $\epsilon$.

The challenge of our task is as follows. Let $\hat{V}_{\epsilon}\hat{U}_{\epsilon} = \hat{U}_{\text{ideal}}$, recovering ideal evolution is equivalent to constructing the compensation unitary $\hat{V}_{\epsilon}=\hat{U}_{\text{ideal}}\hat{U}_{\epsilon}^\dag$. However, direct implementation of $\hat{V}_{\epsilon}$ is not a easy task, partly because the error level $\epsilon$ is \textit{not known} beforehand. Furthermore, even if we could somehow find out the value of $\epsilon$, the fact that $\hat{E}$ does not commute with $\hat{G}$ would prevent us from implementing $\hat{V}_{\epsilon}$ via single qubit rotations, which is still the case even under linear approximation with respect to $\epsilon$, as we will discuss below. Instead, the dependence of $\hat{V}_{\epsilon}$ on $\epsilon$ appears in nested commutators. Series expansion of $\hat{V}_{\epsilon}$ will be given in Appendix.~\ref{app:gbc}. A conceptual sketch of $\hat{U}_{\text{ideal}}$ and $\hat{V}_\epsilon$ is given in Fig.~\ref{Fig_errors}(a).

To obtain a nontrivial error compensation scheme, one should have a better understanding of the error geometry. From the perspective of the Lie group, we can define the generator of the compensation unitary as 
\begin{equation}
\hat{K}_{\mathcal{G}} = -i\frac{\partial}{\partial\epsilon}(\hat{U}_{\text{ideal}}\hat{U}_{\epsilon}^{\dagger})|_{\epsilon = 0},
\end{equation} 
Then, the compensation unitary satisfies $V_{\epsilon} = \exp\left[i\epsilon \hat{K}_{\mathcal{G}} + o(\epsilon^2)\right]$.

Note that the generator $\hat{K}_{\mathcal{G}}$ here is independent of $\epsilon$, and as a function of $\Theta$, $\hat{G}$ and $\hat{E}$. As long as $\{\hat{G}, \hat{E}\} = 0$, a closed form of $\hat{K}_{\mathcal{G}}$ can be given as
\begin{equation}
    \hat{K}_{\mathcal{G}} = \frac{\hat{G}^{-1}}{2i\Theta}\left(e^{2i\Theta\hat{G}}-\mathbb{I}\right)\hat{E}.
\end{equation}
We can actually generalize the generator of compensation unitary to higher-order (Appendix.~\ref{app:gbc}), but the current first-order expression is sufficient for our schemes.

It can be easily checked that the criteria $\{\hat{G},\hat{E}\}=0$ is satisfied in our case, and we have the explicit expression $\hat{K}_{\mathcal{G}} = \frac{2}{\pi}(\mathbb{I}+i\sigma_x^{j_1}\sigma_x^{j_2})(\sigma_z^{j_1}+\sigma_z^{j_2})$. As we claimed earlier, $\hat{K}_{\mathcal{G}}$ is hard to implement directly, yet we can concatenate three gates to cancel the linear contribution of error, by exploiting the fact that the frequency shift error does not change rapidly over time. Keeping the leading term with respect to $\epsilon$ only, we secure the following result:
\begin{equation}
    \hat{U}^{(gbc)}_{\epsilon} = \hat{U}_{\epsilon}\hat{U}_{2\epsilon}^\dagger \hat{U}_{\epsilon} \cong \left(\hat{U}_{\text{ideal}} \hat{U}_\epsilon^{\dagger}\right)\hat{U}_\epsilon = \hat{U}_{\text{ideal}},
\end{equation}
which justifies our {generator-based compensation} scheme. Here we use $\cong$ to denote equivalence by neglecting terms of second order or higher in $\epsilon$. Our GBC scheme is sketch in Fig.~\ref{Fig_tanSpace}(c), and Fig.~\ref{Fig_tanSpace}(a) and (b) provide a heuristic demonstration of the generator analysis on the tangent space. 

\subsection{Quantum circuit implementation of the scheme}
To implement $\hat{U}^{(gbc)}_\epsilon$, the remaining task is to realize $\hat{U}_{2\epsilon}^\dagger = \exp(-i\Theta\hat{G}+i\cdot 2\epsilon\hat{E})$, which is equivalent to an entangling gate with rotation angle $-\Theta$ and error level $-2\epsilon$.
We are able to tune the entangling angle freely by properly modulating the laser pulses (Appendix.~\ref{app:amplitude}). Moreover, the error level  $2\epsilon\equiv2\Delta\omega\tau$ can be achieved by doubling the gate time $\tau\rightarrow2\tau$. So we are able to implement 
$\hat{U}_{2\epsilon}(-\Theta) \equiv \exp\left(-i\Theta\hat{G}-i\cdot 2\epsilon\hat{E}\right)$
which differs from $\hat{U}^\dagger_{2\epsilon}$ by a negative sign on the error term. The error term is inherently determined by the system and cannot be changed directly. However, we can effectively flip its sign by applying additional global $\sigma_x$ gates $\hat{\Pi}\equiv{\sigma}_{x}^{j_1}\otimes{\sigma}_{x}^{j_2}$. Because 
$\hat{\Pi}\hat{G}\hat{\Pi}=\hat{G}$ and $\hat{\Pi}\hat{E}\hat{\Pi}=-\hat{E}$,
the desired gate $\hat{U}_{2\epsilon}^\dagger$ can be obtained by sandwiching $\hat{U}_{2\epsilon}(-\Theta)$ with two $\hat{\Pi}$ gates, i.e.
\begin{eqnarray}
        \hat{U}_{2\epsilon}^\dagger= \hat{\Pi}\hat{U}_{2\epsilon}(-\Theta)\hat{\Pi} . 
\end{eqnarray}
The full quantum circuit implementation our scheme is given below:
$$\Qcircuit @C=1em @R=.7em {
& & & & \hat{U}_{2\epsilon}^{\dagger}(\Theta)  & & & &\\
& & & & & & & &\\
|j_1\rangle && \multigate{1}{\hat{U}_{\epsilon}(\Theta)} & \gate{{\sigma}_x} & \multigate{1}{\hat{U}_{2\epsilon}(-\Theta)} & \gate{{\sigma}_x} &\multigate{1}{\hat{U}_{\epsilon}(\Theta)} & \qw\\
|j_2\rangle && \ghost{\hat{U}_{\epsilon}(\Theta)} & \gate{{\sigma}_x} & \ghost{\hat{U}_{2\epsilon}(-\Theta)} & \gate{{\sigma}_x} & \ghost{\hat{U}_{\epsilon}(\Theta)} & \qw \gategroup{3}{4}{4}{6}{.7em}{--}}$$
where the two-qubit gates $\hat{U}_{\epsilon}(\Theta)$ and $\hat{U}_{2\epsilon}(-\Theta)$ are implemented via amplitude modulation introduced in the previous section, with gatetime $\tau$ and $2\tau$, respectively.\\

\subsection{$X$ gate error and correction} 
\label{subsec:supcode}

Since the asymmetric error stems from the fluctuation of laser frequencies or internal energy levels, single-qubit gates may also be subject to $\sigma_z$-type errors. To quantify this effect, we may assume that the $\hat{\Pi}$ gate is implemented through Rabi oscillation with Rabi frequency $\Omega$. Then a frequency drift of $\Delta\omega$ leads to the effect
\begin{equation}
    \begin{aligned}
        \tilde{\Pi} \hat{G}\tilde{\Pi} - \hat{\Pi} \hat{G}\hat{\Pi}&= o((\Delta\omega/\Omega)^2),\\
        \tilde{\Pi}\hat{E}\tilde{\Pi} - \hat{\Pi} \hat{E}\hat{\Pi}&= 2(\Delta\omega/\Omega) \hat{\Pi} + o((\Delta\omega/\Omega)^2).
    \end{aligned}
\end{equation}
We can therefore define the error level for a single qubit gate as $\epsilon_{\text{sig}}=2(\Delta\omega/\Omega) $. As long as $\Omega \tau >>1$, we have $\epsilon_{\text{sig}}<<\epsilon$. In practice, typical values of $\tau$ and $\Omega$ are at the order of $\tau \sim 10^{-4}$ s, $\Omega \sim 10^6$ Hz, satisfying the assumption above~\cite{zhang2020submicrosecond,grzesiak2020efficient,nakav2023effect}. Therefore, the effect of the error in single-qubit gates appears only in higher-order terms of $\epsilon$. Intuitively, this phenomenon is due to the significantly shorter gatetime of single-qubit gates compared to two-qubit gates.

Alternatively, we could apply the SUPCODE scheme to improve the robustness of single-qubit gates, which exploits the idea of composite pulse~\cite{wang2012composite, brown2005erratum}. During our single qubit gate implementation, the Hamiltonian is in the form of 
$\hat{H}_{\text{sig}}(\Omega_{\text{sig}})=\frac{1}{2}\Omega_{\text{sig}}\sigma_x+\Delta\omega\sigma_z$, where $\Omega_{\text{sig}}\in\left[0,\Omega\right]$ is the Rabi frequency. The simplest 3-piece SUPCODE for single-qubit Pauli-$X$ gate is given by 
\begin{equation}
    \hat{X}_{\text{SUP}} = e^{-i\hat{H}_{\text{sig}}(\Omega)t}e^{-i\hat{H}_{\text{sig}}(\Omega/2)\:2t}e^{-i\hat{H}_{\text{sig}}(\Omega)t},
\end{equation}
where $t=\pi/\Omega$.
It has been verified that the error of $\hat{X}_{\text{SUP}}$ is at the order of $O((\Delta\omega t)^2)$~\cite{wang2012composite}.

\section{\label{sec:numerics} Numerical Simulation}

We evaluate our gate scheme via numerical simulation of two \(\mathrm{^{40}Ca^+}\) ions confined in a potential with a radial trap frequency of $2\pi\times 1.59$ MHz. The equilibrium distance between the ions is \(d = 8\, \mu\text{m}\). We utilize transverse phonon modes to mediate entanglement \cite{zhu2006trapped}, with the center-of-mass mode frequency equals to the radial trap frequency, while the relative motion mode frequency is approximately $2\pi\times 1.48$ MHz. 

\begin{figure}[t]
    \centering
    \includegraphics[width=\linewidth]{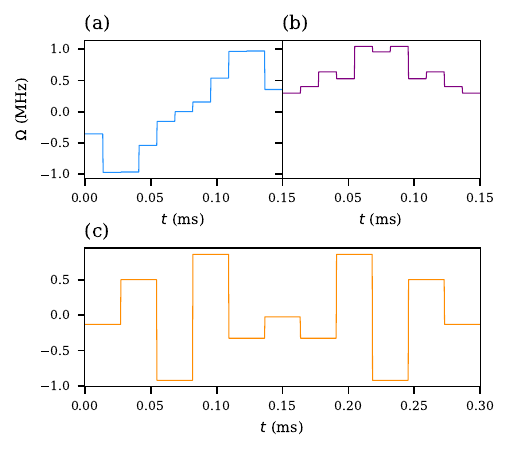}
    \caption{Amplitude modulation waveforms. (a) Non-robust waveform with gate time \(\tau = 150\, \mu\text{s}\) and entangle angle \(\Theta = \frac{\pi}{4}\). (b) symmetry error-robust waveform with gate time \(\tau = 150\, \mu\text{s}\) and entangle angle \(\Theta = \frac{\pi}{4}\). (c) symmetry error-robust waveform with gate time \(\tau = 300\, \mu\text{s}\) and entangle angle \(\Theta = -\frac{\pi}{4}\), which is used in GBC scheme.}
    \label{fig:waveform}
\end{figure}

\begin{figure}[t]
    \centering
    \includegraphics[width=\linewidth]{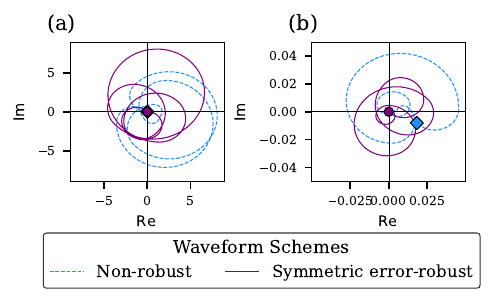}
    \caption{Phase-space trajectories for non-robust waveforms and symmetry error-robust waveform with an asymmetric detuning \(\Delta\omega = 1\, \text{kHz}\). The Lamb-Dicke parameter \(\eta\) is set to be 0.1. (a) Phase-space trajectory for $\alpha_m(t)\eta_m$; (b) Phase-space trajectory for $\Delta\omega\beta_m(t)\eta_m$. }
    \label{fig:pst}
\end{figure}

\begin{figure}[h]
    \centering
    \includegraphics[width=\linewidth]{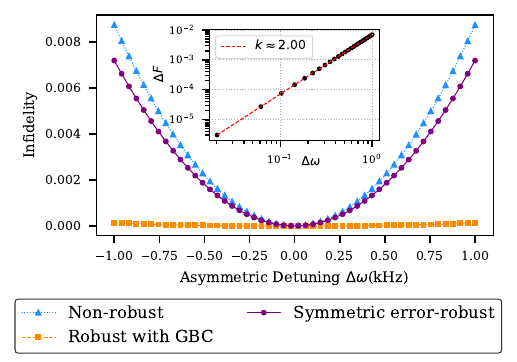}
    \caption{Robustness against asymmetric detuning \(\Delta\omega\) evaluated via numerical simulation for a two-ion \(\mathrm{^{40}Ca^+}\) system. Main panel: Gate infidelity versus asymmetric detuning \(\Delta\omega\). Inset: linear fitting of \(\Delta F\), the difference in fidelity for the symmetry error-robust waveform design and GCB schemes, as a function of \(\Delta\omega\) on a log-log scale. The fitted slope \(k \approx 2\) indicates a quadratic dependency, consistent with Eq.~\eqref{eq:U_tau}. Simulation parameters are provided in the main text.}
    \label{fig:performance_comparison}
\end{figure}
In our numerical simulation, we compare our GBC scheme with two previous schemes. The first scheme is the non-robust waveform, with which we only ensure that the rotation angle is correct, and the spin, and phonon degree of freedoms are decoupled at the end of the operations (i.e. the trajectory of in-phase space ends up at the original point). The second scheme is the one optimized with the symmetry error-robust waveform design. Besides the correctness in the ideal case, it is expected that the trajectory in phase space is still closed in the presence of errors, thus suppressing both the symmetric error and the displacement-dependent part of the asymmetric errors. 

\begin{figure*}
    \centering
    \includegraphics[width=\linewidth]{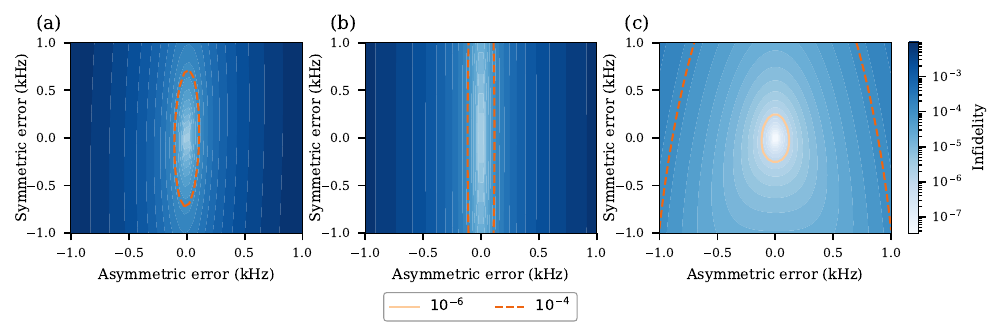}
    \caption{Contour graph of gate infidelity against symmetric and asymmetric errors. (a) Non-robust waveform (b) Symmetry error-robust waveform (c) GBC waveform}
    \label{fig:infid_contour}
\end{figure*}

Following the amplitude modulation approach described in Appendix.~\ref{app:amplitude}, we select a gate time \(\tau = 150\, \mu\text{s}\), bichromatic laser detuning \(\omega_d = 9.470\, \text{MHz}\), and target phase angle \(\Theta = \pi/{4}\) for non-robust and the symmetry error-robust waveform design. 
The GBC scheme contains three pieces of entangling gates and four single-qubit gates. Single qubit gates are assumed to be ideal, and $\hat{U}_{\epsilon}(\Theta)$ is identical to the symmetry error-robust waveform design, and the evolution $\hat{U}^\dag_{2\epsilon}(-\Theta)$ is optimized with bichromatic laser detuning \(\omega_d = 10.190\, \text{MHz}\) and gate time \(\tau = 300\, \mu\text{s}\) to ensure that the error level is doubled. 
The optimized waveforms are illustrated in Fig.~\ref{fig:waveform}.

In Fig.~\ref{fig:pst}, we demonstrate the phase-space trajectory under asymmetric noise for both $\alpha_m(t)$ and $\beta_m(t)$. They correspond to the symmetric error and displacement-dependent asymmetric errors respectively. It can be verified that for non-robust waveform $\alpha_m$ is closed while $\beta_m$ is not. This result is expected because we have only added asymmetric noise in the simulation. On the other hand, for waveform optimized with symmetry error-robust waveform design, the phase-space trajectory for both $\alpha_m(t)$ and $\beta_m(t)$ are closed at the end of the evolution. This is consistent with the fact that the symmetry error-robust waveform design can effectively suppress displacement-dependent asymmetric errors.

To assess the robustness of the schemes, we compare the gate fidelity in the presence of asymmetric detuning errors ranging from \(-10^3\) Hz to \(10^3\) Hz, using three different waveform conditions. Given the coherence time of \(^{40}\mathrm{Ca}^+\) ions' ground state, \(T_2 = 18 \pm 1 \, \mathrm{ms}\) \cite{pogorelov2021compact}, we estimate frequency drifts to be within \(100 \, \mathrm{Hz}\). Yet, we consider a broader range of detuning errors to account for potential imperfections in experimental control and laser miscalibration. As shown in Fig.~\ref{fig:performance_comparison}, our scheme demonstrates significant improvements in fidelity. We analyze the infidelity difference \(\Delta F\) between GBC and the symmetry error-robust waveform schemes. By performing a linear fit on a log-log scale between \(\Delta F\) and the asymmetric detuning \(\Delta\omega\), we find a polynomial index of 2.00. This quadratic scaling behavior is consistent with the error term introduced by asymmetric detuning, \(- \frac{i}{2} \Delta\omega \tau \sum_j \sigma_z^j\) in Eq.~\eqref{eq:U_tau}.

In practice, both symmetric and asymmetric errors are important noise sources. We further investigate the performance with the presence of both noise sources. As shown in Fig.~\ref{fig:infid_contour}(a), for the non-robust waveform, gate infidelity is more sensitive to asymmetric errors compared to the symmetric one. Under a symmetric noise level of 714.3 Hz or asymmetric noise level of 142.8 Hz, the gate infidelity of non-robust waveform increases to above $0.01\%$.
From Fig.~\ref{fig:infid_contour}(b), the symmetry error-robust waveform shows high robustness against symmetric errors, while still sensitive to asymmetric one. An asymmetric error of 138.4 Hz can cause the gate infidelity $ >0.01\%$.
On the other hand, as shown in Fig.~\ref{fig:infid_contour}(c), our scheme shows a significant improvement in robustness against both symmetric and asymmetric errors. The infidelity 0.01\% can be achieved with an arbitrary noise level below 500HZ.

In our simulations above, we have only considered two-ion cases in the presence of asymmetric errors. Yet, our approaches also work for multi-ion systems. Supplementary results for multi-ion cases are provided in Appendix.~\ref{app:sup_num}.




\section{\label{sec:conclusion}Conclusion}

We have tackled the challenge of asymmetric errors in trapped-ion quantum computing. Through comprehensive theoretical and numerical analyses, we demonstrate that our GBC scheme can suppress first-order asymmetric errors, at the same time maintaining the robustness against symmetric errors. Our findings not only enhance our understanding of error dynamics in trapped-ion systems, but provide practical strategies for improving the reliability and performance of quantum computing. Due to the simplicity and generality, our approach allows for direct implementation across various trapped-ion architectures, thereby improving the robustness of trapped-ion-based quantum computing.  Furthermore, the GBC method is general and could be extended to other quantum systems and noise sources with similar geometric properties.

$\\$
\noindent\textbf{Acknowledgement} The authors thank Luming Duan, Xiang Zhang, Zixuan Huo for helpful discussions. This work is supported by the National Natural Science Foundation of China (No.~12405013, No.~12175003 and No.~12361161602), Innovation Program for Quantum Science and Technology (2021ZD0301601), Tsinghua University Dushi program, Open Fund of Key Laboratory of Atomic and Subatomic Structure and Quantum Control (Ministry of Education), NSAF (Grant No.~U2330201)

\bibliographystyle{unsrt}

\onecolumngrid
\newpage
\appendix


\section{Higher Order Terms in Magnus Expansion}
\label{app:higher_mag}




In the main text, we have used Magnus expansion to derive the form of the effective evolution operator, and we retain up to $2$-nd order Magnus terms and neglect higher order contributions. Though this approach is commonly adopted when dealing with the noiseless entangling gate, it lacks a solid validation when noise is presented. In this Appendix we analytically deal with the next order Magnus term and provide a closed form of the effect of the asymmetric error defined in the main text.

\subsection{Single Mode}
As a simplified case, we first consider the following Hamiltonian with only one motional mode:
\begin{equation}
    \begin{aligned}
    \hat{H}_{int}(t)=\lambda\hat{S}_z+\eta\Omega(t)\left( \hat{a}^{\dagger}e^{i\epsilon t}+\hat{a}e^{-i\epsilon t} \right)\hat{S}_{\phi}
\end{aligned}
\end{equation}
here $\hat{S}_{\phi}=\hat{S}_{x}\equiv\frac{1}{2}(\sigma_x^{j_1}+\sigma_x^{j_2})$ is the $x$ direction total spin operator of the two selected ions (labeled by $j_1$ and $j_2$), $\epsilon$ is the laser detuning from the mode frequency, $\eta$ is the Lamb-Dicke parameter and $\lambda$ characterizes the strength of asymmetric error, which crresponds to $\Delta\omega$ in the main text. Using Magnus expansion $\hat{U}=\exp(\sum_i \hat{M}_i)$, the first few Magnus terms are given as:
\begin{equation}
    \begin{aligned}
    \hat{M}_1&=-i\int_0^\tau\hat{H}_{int}(t)\:{\rm d}t\\
    \hat{M}_2&=-\frac{1}{2}\int_0^{\tau}\:{\rm d}t_1\int_0^{t_1}\:{\rm d}t_2\left[\hat{H}_{int}(t_1), \hat{H}_{int}(t_2)\right]\\
    \hat{M}_3&=\frac{i}{6}\int_0^{\tau}\:{\rm d}t_1\int_0^{t_1}\:{\rm d}t_2\int_0^{t_2}\:{\rm d}t_3\left(\left[\hat{H}_{int}(t_1), \left[\hat{H}_{int}(t_2), \hat{H}_{int}(t_3)\right]\right]+\left[\hat{H}_{int}(t_3), \left[\hat{H}_{int}(t_2), \hat{H}_{int}(t_1)\right]\right]\right)
\end{aligned}
\end{equation}
The first order term can be easily obtained:
\begin{equation}
    \begin{aligned}
    \hat{M}_1&=-i\left[\lambda\hat{S}_z\tau+\eta\left(\alpha(\tau)\hat{a}^\dagger+\alpha(\tau)^*\hat{a}\right)\hat{S}_x\right]
\end{aligned}
\end{equation}
where $\alpha(\tau)$ is defined as
\begin{equation}
    \alpha(\tau)=\int_0^{\tau}\Omega(t)e^{i\epsilon t}\:{\rm d}t
\end{equation}
Within the amplitude modulation (AM) scheme, we optimize the Rabi frequency $\Omega(t)$ of the laser pulse such that $\alpha(\tau)=0$, hence the first order term reduces to 
\begin{equation}
    \hat{M}_1=-i\lambda\hat{S}_z\tau
\end{equation}
The derivation of the second order term is also straightforward:
\begin{equation}
    \begin{aligned}
    \hat{M}_2&=-\frac{1}{2}\int_0^\tau\:{\rm d}t_1\int_0^{t_1}\:{\rm d}t_2\left\{\left[-\eta\Omega(t_1)\left( \hat{a}^{\dagger}e^{i\epsilon t_1}+\hat{a}e^{-i\epsilon t_1} \right)\hat{S}_x,-\eta\Omega(t_2)\left( \hat{a}^{\dagger}e^{i\epsilon t_2}+\hat{a}e^{-i\epsilon t_2} \right)\hat{S}_x\right]\right. \\
    &- \left.\eta\lambda\Omega(t_1)\left( \hat{a}^{\dagger}e^{i\epsilon t_1}+\hat{a}e^{-i\epsilon t_1} \right)\left[\hat{S}_x,\hat{S}_z\right]-\eta\lambda\Omega(t_2)\left( \hat{a}^{\dagger}e^{i\epsilon t_2}+\hat{a}e^{-i\epsilon t_2} \right)\left[\hat{S}_z,\hat{S}_x\right]\right\}
\end{aligned}
\end{equation}
where the first term has the form of $i\cdot 2\Theta\hat{S}_x^2$, which is equivalent to $i\Theta\sigma_x^{j_1}\sigma_x^{j_2}$ up to a global phase. Here 
\begin{equation}
    \Theta=-\frac{1}{2}{\rm Im}\left[\eta^2\int_0^{\tau}\:{\rm d}t_1\int_0^{t_1}\:{\rm d}t_2\:\Omega(t_1)\Omega(t_2)e^{-i\epsilon(t_1-t_2)}\right]
\end{equation}
is the entangling angle. In our AM scheme setting, we optimize the pulse to achieve $\Theta=\pi/4$ for a maximum entangling gate. The remaining terms write:
\newpage
\begin{equation}
    \begin{aligned}
    &\quad-\frac{1}{2}\int_0^\tau\:{\rm d}t_1\int_0^{t_1}\:{\rm d}t_2\left\{-\eta\lambda\Omega(t_1)\left( \hat{a}^{\dagger}e^{i\epsilon t_1}+\hat{a}e^{-i\epsilon t_1} \right)+\eta\lambda\Omega(t_2)\left( \hat{a}^{\dagger}e^{i\epsilon t_2}+\hat{a}e^{-i\epsilon t_2} \right)\right\}\left[\hat{S}_x, \hat{S}_z\right]\\
    &=-\frac{\lambda}{2}\hat{S}_y\left[i\eta\int_0^{\tau}\:{\rm d}t_1\int_0^{t_1}\:{\rm d}t_2\:\Omega(t_1)\left( \hat{a}^{\dagger}e^{i\epsilon t_1}+\hat{a}e^{-i\epsilon t_1} \right) - i\eta\int_0^{\tau}\:{\rm d}t_1\int_0^{t_1}\:{\rm d}t_2\:\Omega(t_2)\left( \hat{a}^{\dagger}e^{i\epsilon t_2}+\hat{a}e^{-i\epsilon t_2} \right)\right]\\
    &=-\frac{\lambda}{2}\hat{S}_y\left\{\left[\frac{\partial\alpha}{\partial\epsilon}(\tau)-i\int_0^{\tau}\alpha(t)\:{\rm d}t\right]\hat{a}^{\dagger}-h.c.\right\}
\end{aligned}
\end{equation}
Actually, using a robust AM scheme we can simultaneously eliminate both terms in the bracket here, with a symmetric pulse shape $\Omega(t)\equiv\Omega(\tau-t)$. It has been proved that for such a pulse, $\frac{\partial\alpha}{\partial\epsilon}(\tau)=0$ as long as $\int_0^{\tau}\alpha(t)\:{\rm d}t=0$. Consequently, we have secured the second order term
\begin{equation}
    \hat{M}_2 = i\cdot 2\Theta\hat{S}_x^2.
\end{equation}
Now we turn to the third order term in the Magnus expansion. We do the integration and obtain
\begin{equation}
    \begin{aligned}
    &\int_0^{\tau}\:{\rm d}t_1\int_0^{t_1}\:{\rm d}t_2\int_0^{t_2}\:{\rm d}t_3 \left(\left[\hat{H}_{int}(t_1), \left[\hat{H}_{int}(t_2), \hat{H}_{int}(t_3)\right]\right]+\left[\hat{H}_{int}(t_3), \left[\hat{H}_{int}(t_2), \hat{H}_{int}(t_1)\right]\right]\right)\\
    &\qquad= A+B+C+D
\end{aligned}
\end{equation}
with $A, B, C, D$ defined as
\begin{equation}
    \begin{aligned}
    A&= \left[\lambda\hat{S}_z, \int_0^{\tau}\:{\rm d}t_1\int_0^{t_1}\:{\rm d}t_2\int_0^{t_2}\:{\rm d}t_3 \left[\hat{H}_{int}(t_2), \hat{H}_{int}(t_3)\right]\right]\\
    B&= \int_0^\tau\:{\rm d}t_1\left[-\eta\Omega(t_1)(\hat{a}^{\dagger}e^{i\epsilon t_1} + \hat{a}e^{-i\epsilon t_1})\hat{S}_x, \int_0^{t_1}\:{\rm d}t_2\int_0^{t_2}\:{\rm d}t_3 \left[\hat{H}_{int}(t_2), \hat{H}_{int}(t_3)\right]\right]\\
    C&= \left[\lambda\hat{S}_z, \int_0^{\tau}\:{\rm d}t_1\int_0^{t_1}\:{\rm d}t_2\int_0^{t_2}\:{\rm d}t_3\left[\hat{H}_{int}(t_2), \hat{H}_{int}(t_1)\right]\right]\\
    D&= \int_0^{\tau}\:{\rm d}t_1\int_0^{t_1}\:{\rm d}t_2\int_0^{t_2}\:{\rm d}t_3 \left[ -\eta\Omega(t_3)(\hat{a}^{\dagger}e^{i\epsilon t_3}+\hat{a}e^{-i\epsilon t_3})\hat{S}_x, \left[\hat{H}_{int}(t_2), \hat{H}_{int}(t_1)\right]\right]
\end{aligned}
\end{equation}
Note that the term $\int_0^{t_1}\:{\rm d}t_2\int_0^{t_2}\:{\rm d}t_3 \left[\hat{H}_{int}(t_2), \hat{H}_{int}(t_3)\right]$ appears in the intermediate step of computing $\hat{M}_2$, and we may plug in the expression for $\hat{M}_2$, except that all instances of $\tau$ in $\hat{M}_2$ are replaced by $t_1$:
\begin{equation}
    \hat{M}_2(t_1) = i\cdot 2\Theta(t_1)\hat{S}_x^2 - \frac{\lambda\eta}{2}\hat{S}_y\left\{\left[\frac{\partial\alpha}{\partial\epsilon}(t_1) -i \int_0^{t_1}\alpha(t)\:{\rm d} t\right]\hat{a}^{\dagger}-h.c.\right\}
\end{equation}
where $\Theta(t)$ is defined as 
\begin{equation}
    \Theta(t)=-\frac{1}{2}{\rm Im}\left[\eta^2\int_0^{t}\:{\rm d}t_1\int_0^{t_1}\:{\rm d}t_2\:\Omega(t_1)\Omega(t_2)e^{-i\epsilon(t_1-t_2)}\right]
\end{equation}
and $\Theta(\tau)\equiv \Theta$ reduces to $\pi/4$. Using this formula, we are ready to compute the four terms of $A$, $B$, $C$ and $D$:
\newpage
\begin{equation}
    \begin{aligned}
    A&=-4i\lambda\int_0^{\tau}\Theta(t_1)\:{\rm d}t_1\left[\hat{S}_z, \hat{S}_x^2\right]-i\lambda^2\eta\hat{S}_x\int_0^{\tau}\:{\rm d}t_1\left\{\left[\frac{\partial\alpha}{\partial\epsilon}(t_1) -i \int_0^{t_1}\alpha(t)\:{\rm d} t\right]\hat{a}^{\dagger}-h.c.\right\}\\
    B&=\lambda\eta^2\int_0^{\tau}\:{\rm d}t_1\:\Omega(t_1)e^{i\epsilon t_1}\left[\hat{a}^{\dagger}\hat{S}_x,-\left\{\left[\frac{\partial\alpha}{\partial\epsilon}(t_1) -i \int_0^{t_1}\alpha(t)\:{\rm d} t\right]\hat{a}^{\dagger}-h.c.\right\}\hat{S}_y\right]\\
    &+\lambda\eta^2\int_0^{\tau}\:{\rm d}t_1\:\Omega(t_1)e^{-i\epsilon t_1}\left[\hat{a}\hat{S}_x,-\left\{\left[\frac{\partial\alpha}{\partial\epsilon}(t_1) -i \int_0^{t_1}\alpha(t)\:{\rm d} t\right]\hat{a}^{\dagger}-h.c.\right\}\hat{S}_y\right]\\
    &=2{\rm Re}\left(\lambda\eta^2\int_0^{\tau}\:{\rm d}t_1\:\Omega(t_1)e^{i\epsilon t_1}\left[\hat{a}^{\dagger}\hat{S}_x,-\left\{\left[\frac{\partial\alpha}{\partial\epsilon}(t_1) -i \int_0^{t_1}\alpha(t)\:{\rm d} t\right]\hat{a}^{\dagger}-h.c.\right\}\hat{S}_y\right]\right)\\
    C&=\lambda\eta^2\left[\hat{S}_z, \hat{S}_x^2\right]\int_0^{\tau}\:{\rm d}t_1\int_0^{t_1}\:{\rm d}t_2\:\Omega(t_1)\Omega(t_2)\left[e^{-i\epsilon(t_2-t_1)}-e^{i\epsilon(t_2-t_1)}\right]t_2\\
    &- \lambda^2\eta\hat{S}_x\left\{\int_0^{\tau}\:{\rm d}t_1\cdot \frac{1}{2}t_1^2\Omega(t_1)\left(\hat{a}^{\dagger}e^{i\epsilon t_1}+\hat{a}e^{-i\epsilon t_1}\right) - \int_0^{\tau}\:{\rm d}t_1\int_0^{t_1}\:{\rm d}t_2\cdot t_2\Omega(t_2)\left(\hat{a}^{\dagger}e^{i\epsilon t_2}+\hat{a}e^{-i\epsilon t_2}\right)\right\}\\
    D&=i\lambda\eta^2\int_0^{\tau}\:{\rm d}t_1\int_0^{t_1}\:{\rm d}t_2\int_0^{t_2}\:{\rm d}t_3 \:\Omega(t_3)\left\{\Omega(t_1)\left[(\hat{a}^{\dagger}e^{i\epsilon t_3}+\hat{a}e^{-i\epsilon t_3})\hat{S}_x,  (\hat{a}^{\dagger}e^{i\epsilon t_1}+\hat{a}e^{-i\epsilon t_1})\hat{S}_y\right]\right.\\
    &-\left.\Omega(t_2)\left[(\hat{a}^{\dagger}e^{i\epsilon t_3}+\hat{a}e^{-i\epsilon t_3})\hat{S}_x,  (\hat{a}^{\dagger}e^{i\epsilon t_2}+\hat{a}e^{-i\epsilon t_2})\hat{S}_y\right]\right\}
\end{aligned}
\end{equation}
Our task then is to simplify these integrals. Notice that for those terms whose $\hat{a}^{\dagger}$ and $\hat{a}$ factors do not pair up (for instance, terms proportional to $\hat{a}^3\hat{a}^{\dagger}$) have to pair with another term to avoid being canceled up once we partial trace the phonon modes in the evolution operator, thus effectively they appear only by higher orders in our results \cite{wu2018noise}. For our purpose only we will retain those terms with paired up $\hat{a}$ and $\hat{a}^{\dagger}$. Furthermore, these integrals appear in the order of $\lambda^2\eta$ or $\lambda\eta^2$, between which the latter are of greater interest, because we have assumed $\lambda$ is sufficiently small. Keeping relevant terms only yields:
\begin{equation}
    \begin{aligned}
    A&=-4i\lambda\int_0^{\tau}\Theta(t_1)\:{\rm d}t_1\left[\hat{S}_z, \hat{S}_x^2\right]+o(\lambda^2\eta)\\
    B&\cong 2\lambda\eta^2\:{\rm Re}\left\{\int_0^{\tau}\:{\rm d}t_1\:\Omega(t_1)e^{i\epsilon t_1}\left(\frac{\partial\alpha^*}{\partial\epsilon}(t_1)+i \int_0^{t_1}\alpha^*(t)\:{\rm d}t\right)\cdot\left[\hat{a}^{\dagger}\hat{S}_x, \hat{a}\hat{S}_y\right]\right\}+o(\lambda^2\eta)\\
    C&=\lambda\eta^2\left[\hat{S}_z, \hat{S}_x^2\right]\int_0^{\tau}\:{\rm d}t_1\int_0^{t_1}\:{\rm d}t_2\:\Omega(t_1)\Omega(t_2)\left[e^{-i\epsilon(t_2-t_1)}-e^{i\epsilon(t_2-t_1)}\right]t_2+o(\lambda^2\eta)\\
    D&\cong-2\lambda\eta^2\:{\rm Im}\left\{\left[\int_0^{\tau}\:{\rm d}t_1\:\Omega(t_1)e^{-i\epsilon t_1}\int_0^{t_1}\alpha(t)\:{\rm d}t +\int_0^{\tau} \:{\rm d}t_1\int_0^{t_1}\Omega(t)e^{-i\epsilon t}\alpha(t)\:{\rm d}t\right]\cdot\left[\hat{a}^{\dagger}\hat{S}_x, \hat{a}\hat{S}_y\right]\right\}+o(\lambda^2\eta)
\end{aligned}
\end{equation}
To simplify those terms, we consider first the sum of $A+C$, dropping all terms in order $o(\lambda^2\eta)$. Plugging in the definition of $\Theta(t)$ and $\alpha(t)$ and taking $t_2=\int_0^{t_2}\:{\rm d}t$, we obtain
\begin{equation}
    \begin{aligned}
        A& = 2i\lambda\eta^2\:{\rm Im}\left[\int_0^{\tau}\:{\rm d}t\int_0^{t}\:{\rm d}t_1\int_0^{t_1}\:{\rm d}t_2\:\Omega(t_1)\Omega(t_2)e^{-i\epsilon(t_1-t_2)}\right]\cdot\left[\hat{S}_z,\hat{S}^2_x\right]\\
        C& =-2i\lambda\eta^2\:{\rm Im}\left[\int_0^{\tau}\:{\rm d}t_1\int_0^{t_1}\:{\rm d}t_2\int_0^{t_2}\:{\rm d}t\:\Omega(t_1)\Omega(t_2)e^{-i\epsilon(t_1-t_2)}\right]\cdot\left[\hat{S}_z,\hat{S}^2_x\right]
    \end{aligned}
\end{equation}
Note that the integration in $C$ with respect to $t_1$, $t_2$ and $t$ could be reordered via
\begin{equation}
    \int_0^{\tau}\:{\rm d}t_1\int_0^{t_1}\:{\rm d}t_2\int_0^{t_2}\:{\rm d}t\:\Omega(t_1)\Omega(t_2)e^{-i\epsilon(t_1-t_2)} = \int_0^{\tau}\:{\rm d}t\int_t^{\tau}\:{\rm d}t_2\int_{t_2}^{\tau}\:{\rm d}t_1\:\Omega(t_1)\Omega(t_2)e^{-i\epsilon(t_1-t_2)},
\end{equation}
since they both do integration over the $3$-dim region defined by $0\leqslant t\leqslant t_2\leqslant t_1\leqslant \tau$. Further define $t'_1 = \tau-t_1$, $t'_2 = \tau - t_2$ and $t' = \tau - t$, this integration becomes
\begin{equation}
\begin{aligned}
    \int_0^{\tau}\:{\rm d}t\int_t^{\tau}\:{\rm d}t_2\int_{t_2}^{\tau}\:{\rm d}t_1\:\Omega(t_1)\Omega(t_2)e^{-i\epsilon(t_1-t_2)}&=\int_0^{\tau}\:{\rm d}t'\int_0^{t'}\:{\rm d}t'_2\int_{0}^{t'_2}\:{\rm d}t'_1\:\Omega(\tau - t'_1)\Omega(\tau - t'_2)e^{-i\epsilon(t'_2-t'_1)}\\
    &=\int_0^{\tau}\:{\rm d}t\int_0^{t}\:{\rm d}t_1\int_{0}^{t_1}\:{\rm d}t_1\:\Omega(t_2)\Omega(t_1)e^{-i\epsilon(t_1-t_2)} 
\end{aligned}
\end{equation}
where the last step is achieved by relabeling $t'\rightarrow t$, $t'_1\rightarrow t_2$ and $t'_2\rightarrow t_1$, and exploiting the symmetry property $\Omega(\tau - t)\equiv\Omega(t)$. As a result, we secure 
\begin{equation}
    A+C = 0
\end{equation}
accurate up to leading order. The leading order expressions of $B$ and $D$ can also be expended as
\begin{equation}
\begin{aligned}
    B&=2\lambda\eta^2\:{\rm Im}\left(I_B\cdot\left[\hat{a}^{\dagger}\hat{S}_x, \hat{a}\hat{S}_y\right]\right)\\
    D&=-2\lambda\eta^2\:{\rm Im}\left(I_D\cdot\left[\hat{a}^{\dagger}\hat{S}_x, \hat{a}\hat{S}_y\right]\right)
\end{aligned}
\end{equation}
where 
\begin{equation}
    \begin{aligned}
        I_B& = \int_0^{\tau}{\rm d}t_1 \int_0^{t_1}{\rm d}t_2 \int_0^{t_2}{\rm d}t\:\Omega(t_1)\Omega(t_2)e^{i\epsilon(t_1-t_2)} - \int_0^{\tau}{\rm d}t_1 \int_0^{t_1}{\rm d}t_2 \int_0^{t_2}{\rm d}t\:\Omega(t_1)\Omega(t)e^{i\epsilon(t_1-t)}\\
        I_D& = \int_0^{\tau}{\rm d}t_1 \int_0^{t_1}{\rm d}t \int_0^{t}{\rm d}t_2 \:\Omega(t_1)\Omega(t_2)e^{-i\epsilon(t_1-t_2)} + \int_0^{\tau}{\rm d}t_1 \int_0^{t_1}{\rm d}t \int_0^{t}{\rm d}t_2 \:\Omega(t)\Omega(t_2)e^{-i\epsilon(t-t_2)}
    \end{aligned}
\end{equation}
By relabeling $t\leftrightarrow t_2$ and re-ordering the integration $\int_0^{t_1}\:{\rm d}t \int_0^{t}\:{\rm d}t_2\rightarrow \int_0^{t_1}{\rm d}t_2 \int_{t_2}^{t_1}{\rm d}t$, we obtain
\begin{equation}
    I_B = \int_0^{\tau}{\rm d}t_1 \int_0^{t_1}{\rm d}t_2\: (2t_2 - t_1)\:\Omega(t_1)\Omega(t_2)e^{i\epsilon(t_1-t_2)}.
\end{equation}
Similarly, for $I_D$ we have
\begin{equation}
    \begin{aligned}
        I_D &= \int_0^{\tau}{\rm d}t_1 \int_0^{t_1}{\rm d}t_2\: (\tau - t_2)\:\Omega(t_1)\Omega(t_2)e^{-i\epsilon(t_1-t_2)}\\
        &=\int_0^{\tau}{\rm d}t_1\int_{t_1}^{\tau}{\rm d}t_2 \:t_2\:\Omega(t_1)\Omega(t_2)e^{i\epsilon(t_1-t_2)}\\
        &=\int_0^{\tau}{\rm d}t_1\left( \int_0^{\tau}{\rm d}t_2 - \int_0^{t_1}{\rm d}t_2 \right) t_2 \:\Omega(t_1)\Omega(t_2)e^{i\epsilon(t_1-t_2)}
    \end{aligned}
\end{equation}
with the penultimate step derived via changing variable $t'_1=\tau - t_1$, $t'_2 = \tau-t_2$ followed by relabeling. We realize that in the first integral here, $t_1$ and $t_2$ are decoupled, and the integration with respect to $t_1$ gives
\begin{equation}
    \int_0^{\tau}{\rm d}t_1\:\Omega(t_1) e^{i\epsilon t_1} = \alpha(\tau) = 0
\end{equation}
according to our amplitude modulation scheme. Thus we obtain $B+D$ as
\begin{equation}
\begin{aligned}
    B+D &= 2\lambda\eta^2\:{\rm Im}\left((I_B-I_D)\cdot \left[\hat{a}^{\dagger}\hat{S}_x, \hat{a}\hat{S}_y\right]\right)\\
    & = 2\lambda\eta^2\:{\rm Im}\left( \int_0^{\tau}{\rm d}t_1 \int_0^{t_1}{\rm d}t_2\: (3t_2 - t_1)\:\Omega(t_1)\Omega(t_2)e^{i\epsilon(t_1-t_2)}  \cdot \left[\hat{a}^{\dagger}\hat{S}_x, \hat{a}\hat{S}_y\right]\right).
\end{aligned}
\end{equation}
Finally we arrive at the $3$-rd order Magnus term
\begin{equation}
    \begin{aligned}
    \hat{M}_3&=\frac{i}{6}(A+B+C+D)\\
    &=\frac{i}{3}\lambda\eta^2\: {\rm Im}\left\{\int_0^\tau{\rm d}t_1\int_0^{t_1}{\rm d}t_2\:(3t_2-t_1)\Omega(t_1)\Omega(t_2)e^{i\epsilon(t_1-t_2)}\cdot\left[\hat{a}^{\dagger}\hat{S}_x, \hat{a}\hat{S}_y\right]\right\}+o(\lambda^2\eta).
\end{aligned}
\end{equation}
\newpage
\subsection{Multiple modes}
Our next step is to generalize the previous result to multiple mode cases. For the trapped-ion system with $N$ collective vibrational modes, the interaction Hamiltonian now becomes
\begin{equation}
    \hat{H}_{int} = \lambda\hat{S}_z+\Omega(t)\sum_m\eta_m\left(\hat{a}_m^{\dagger}e^{i\delta_m t}+\hat{a}_me^{-i\delta_m t}\right)\hat{S}_x^{(m)}
\end{equation}
By defining 
\begin{equation}
    \begin{aligned}
    \hat{S}_x^{(m)}&=b_{j_1}^m\hat{S}_{1x}+b_{j_2}^m\hat{S}_{2x}\\
    \hat{S}_y^{(m)}&=b_{j_1}^m\hat{S}_{1y}+b_{j_2}^m\hat{S}_{2y}\\
\end{aligned}
\end{equation}
associated with the commutation relation
\begin{equation}
    \begin{aligned}
    [\hat{S}_z, \hat{S}_x^{(m)}]&=i\hat{S}_y^{(m)}\\
    [\hat{S}_z, \hat{S}_y^{(m)}]&=-i\hat{S}_x^{(m)}
\end{aligned}
\end{equation}
Now the first order term becomes
\begin{equation}
    \begin{aligned}
    \hat{M}_1&=-i\left[\lambda\hat{S}_z\tau+\sum_m\:\eta_m\left(\alpha_m(\tau)\hat{a}_m^\dagger+\alpha_m(\tau)^*\hat{a}_m\right)\hat{S}_x^{(m)}\right]
\end{aligned}
\end{equation}
with $\alpha_m$ corresponding to each phonon mode $m$ defined as
\begin{equation}
    \alpha_m(\tau)=\int_0^{\tau}\Omega(t)e^{i\delta_m t}\:{\rm d}t
\end{equation}
and will be tuned to $0$ in an ideal MS gate. The second order term becomes
\begin{equation}
    \begin{aligned}
    \hat{M}_2&=-\frac{1}{2}\sum_{m, m'}\int_0^\tau\:{\rm d}t_1\int_0^{t_1}\:{\rm d}t_2\left[-\eta_m\Omega(t_1)\left( \hat{a}_m^{\dagger}e^{i\delta_m t_1}+\hat{a}_me^{-i\delta_m t_1} \right)\hat{S}_x^{(m)},-\eta_{m'}\Omega(t_2)\left( \hat{a}_{m'}^{\dagger}e^{i\delta_{m'} t_2}+\hat{a}_{m'}e^{-i\delta_{m'} t_2} \right)\hat{S}_x^{(m')}\right] \\
    &+\frac{1}{2}\sum_{m}\int_0^\tau\:{\rm d}t_1\int_0^{t_1}\:{\rm d}t_2 \left\{\eta_m\lambda\left[\Omega(t_2)\left( \hat{a}_m^{\dagger}e^{i\delta_m t_2}+\hat{a}_me^{-i\delta_m t_2} \right)-\Omega(t_1)\left( \hat{a}_m^{\dagger}e^{i\delta_m t_1}+\hat{a}_me^{-i\delta_m t_1} \right)\right]\cdot\left[\hat{S}_z,\hat{S}^{(m)}_x\right]\right\}\\
    &=4i\Theta\hat{S}_{1x}\hat{S}_{2x}-\frac{\lambda}{2}\sum_m\:\eta_m\hat{S}_y^{(m)}\left\{\left[\frac{\partial\alpha_m}{\partial\delta_m}(\tau)-i\int_0^{\tau}\alpha_m(t)\:{\rm d}t\right]\hat{a}_m^{\dagger} - h.c.\right\}
\end{aligned}
\end{equation}
which, under our robust pulse design, simplifies to $2i\Theta\hat{S}_{1x}\hat{S}_{2x}$. Here the entangling angle $\Theta=\Theta(\tau)$ is defined as
\begin{equation}
\Theta(\tau)=-\frac{1}{2}{\rm Im}\left[\sum_m\eta_m^2 b_{j_1}^m b_{j_2}^m\int_0^{\tau}\:{\rm d}t_1\int_0^{t_1}\:{\rm d}t_2\:\Omega(t_1)\Omega(t_2)e^{-i\delta_m(t_1-t_2)}\right]    
\end{equation}
whose value is set to $\pi/4$. For the third order term in Magnus expansion, we follow the definition of $A$, $B$, $C$ and $D$, resulting in 
\begin{equation}
    \begin{aligned}
    A&=-4i\lambda\int_0^{\tau}\Theta(t_1)\:{\rm d}t_1\left[\hat{S}_z, \hat{S}_{1x}\hat{S}_{2x}\right]+o(\lambda^2\eta)\\
    B&\cong 2\lambda\sum_m\eta^2_m\:{\rm Re}\left\{\int_0^{\tau}\:{\rm d}t_1\:\Omega(t_1)e^{i\delta_m t_1}\left(\frac{\partial\alpha_m^*}{\partial\delta_m}(t_1)+i\int_0^{t_1}\alpha_m^*(t)\:{\rm d}t\right)\cdot\left[\hat{a}^{\dagger}_m\hat{S}^{(m)}_x, \hat{a}_m\hat{S}^{(m)}_y\right]\right\}+o(\lambda^2\eta)\\
    C&=2\lambda\left[\hat{S}_z, \hat{S}_{1x}\hat{S}_{2x}\right]\sum_m\eta_m^2 b_{j_1}^m b_{j_2}^m\int_0^{\tau}\:{\rm d}t_1\int_0^{t_1}\:{\rm d}t_2\:\Omega(t_1)\Omega(t_2)\left[e^{-i\delta_m(t_2-t_1)}-e^{i\delta_m(t_2-t_1)}\right]t_2+o(\lambda^2\eta)\\
    D&\cong-2\lambda\sum_m\eta^2_m\:{\rm Im}\left\{\left[\int_0^{\tau}\:{\rm d}t_1\:\Omega(t_1)e^{-i\delta_m t_1}\int_0^{t_1}\alpha_m(t)\:{\rm d}t +\int_0^{\tau} \:{\rm d}t_1\int_0^{t_1}\Omega(t)e^{-i\delta_m t}\alpha_m(t)\:{\rm d}t\right]\cdot\left[\hat{a}^{\dagger}_m\hat{S}^{(m)}_x, \hat{a}_m\hat{S}^{(m)}_y\right]\right\}\\
    &\qquad\qquad\qquad\qquad\qquad\qquad\qquad\qquad\qquad\qquad\qquad\qquad\qquad\qquad\qquad\qquad\qquad\qquad\qquad\:+o(\lambda^2\eta)
    \end{aligned}
\end{equation}

Following analogous calculation procedure to the single mode case, we safely obtain the result:
\begin{equation}
    \hat{M}_3 = \frac{i}{3}\lambda\sum_m\eta^2_m\:{\rm Im}\left\{\int_0^\tau\:{\rm d}t_1\int_0^{t_1}\:{\rm d}t_2\:(3t_2-t_1)\Omega(t_1)\Omega(t_2)e^{i\delta_m(t_1-t_2)}\cdot\left[\hat{a}^{\dagger}_m\hat{S}^{(m)}_x, \hat{a}_m\hat{S}^{(m)}_y\right]\right\} + o(\lambda^2\eta).
    \label{eq:3rd_Res}
\end{equation}
Though actually the high order effect does not pose a great problem, as can be seen in the numerical results in the main text, one may still wonder what could we do when accuracy requirement becomes more strict. Here we point out that in principle, we may impose additional optimization constraint to suppress the higher order effect according to the result Eq.~\ref{eq:3rd_Res}.

\section{Details of generator-based compensation (GBC) scheme}
\label{app:gbc}
\subsection{Framework of GBC scheme}
We start with a general situation, where we want to recover an ideal unitary operator $\hat{U}_{\rm ideal}$ from noisy available unitary operators $\hat{U}_{\epsilon}$, defined as
\begin{equation}
    \begin{aligned}
        &\hat{U}_{\text{ideal}}=\exp\left(i\Theta\hat{G}\right),\\
        &\hat{U}_{\epsilon} = \exp\left(i\Theta\hat{G} - i\epsilon\hat{E}\right),
    \end{aligned}
\end{equation}
Our goal is to find an error-mitigating operator $\hat{V}_{\epsilon}$ which helps fix $\hat{U}_{\epsilon}$ into $\hat{U}_{\text{ideal}}$. Obviously, by requiring $\hat{V}_{\epsilon}\hat{U}_{\epsilon} = \hat{U}_{\text{ideal}}$, we may apply Baker-Campbell-Hausdorff (BCH) formula,
\begin{equation}
    \begin{aligned}
        \ln(e^Ae^B) &= A + B + \frac{1}{2}[A,B] + \frac{1}{12}([A,[A,B]]+[B,[B,A]])-\frac{1}{24}[B,[A,[A,B]]]\\
        &\qquad-\frac{1}{720}([B,[B,[B,[B,A]]]] + [A,[A,[A,[A,B]]]])\\
        &\qquad +\frac{1}{360}([A,[B,[B,[B,A]]]] + [B,[A,[A,[A,B]]]])\\
        &\qquad + \frac{1}{120}([B,[A,[B,[A,B]]]] + [A,[B,[A,[B,A]]]]) + \cdots
    \end{aligned}\label{eq:BCH}
\end{equation}
to compute the series form of $\ln\hat{V}_{\epsilon}$,
\begin{equation}
\begin{aligned}
    \hat{V}_{\epsilon} &= \hat{U}_{\text{ideal}}\hat{U}_{\epsilon}^{\dagger}\\
    & = e^{i\Theta\hat{G}}e^{-i\Theta\hat{G}+i\epsilon\hat{E}}\\
    & = \exp\left\{i\Theta\hat{G} + (-i\Theta\hat{G}+i\epsilon\hat{E}) + \frac{1}{2}\left[i\Theta\hat{G}, i\epsilon\hat{E}\right]\right.\\
    & \qquad+ \left.\frac{1}{12}\left[i\Theta\hat{G},\left[i\Theta\hat{G}, i\epsilon\hat{E}\right]\right] - \frac{1}{12}\left[-i\Theta\hat{G}+i\epsilon\hat{E},\left[i\Theta\hat{G}, i\epsilon\hat{E}\right]\right]+ \cdots\right\}
\end{aligned}\label{eq:series_V}
\end{equation}
where we have simplified the formula by using $\left[i\Theta\hat{G}, -i\Theta\hat{G}+i\epsilon\hat{E}\right] = \left[i\Theta\hat{G}, i\epsilon\hat{E}\right]$. By assuming $\epsilon$ to be sufficiently small, we may safely retain all linear terms of $\epsilon$ in $\ln\hat{V}_{\epsilon}$ and discard higher order terms. However, as we may notice from Eq.\ref{eq:BCH} and Eq.\ref{eq:series_V}, linear terms of $\epsilon$ appear in arbitrarily high order terms of BCH expansion, hence it is difficult to sum them up altogether. But we could directly write down the closed form as
\begin{equation}
    \hat{V}_{\epsilon} = \exp\left(i\epsilon\hat{K}_{\mathcal{G}} + o(\epsilon^2)\right)
    \label{eq:V_epsilon}
\end{equation}
since we know the leading term in the index should be linear in $\epsilon$. Note that $\hat{K}_{\mathcal{G}}$ is the generator of $\hat{V}_{\epsilon}$ in the unitary operator space $SU(4)$, and takes the form of 
\begin{equation}
    \begin{aligned}
        \hat{K}_{\mathcal{G}} &= -i\frac{\partial}{\partial\epsilon}\hat{V}_{\epsilon}\big|_{\epsilon = 0}\\
        & = \hat{U}_{\text{ideal}}\cdot(-i\frac{\partial}{\partial\epsilon}\hat{U}^{\dagger}_{\epsilon}\big|_{\epsilon = 0})
    \end{aligned}
\end{equation}
Generally, $\hat{K}_{\mathcal{G}}$ is linearly independent of $\hat{G}$ or $\hat{E}$, and is not even in the proximity of either operator. Hence it cannot be approximated by $\hat{G}$ or $\hat{E}$ alone. Returning to Eq.~\eqref{eq:V_epsilon}, doubling the value of $\epsilon$ gives
\begin{equation}
    \begin{aligned}
        \hat{V}_{2\epsilon} = \hat{U}_{\text{ideal}}\hat{U}_{2\epsilon}^{\dagger} &= \exp\left(2i\epsilon\hat{K}_{\mathcal{G}} + o(\epsilon^2)\right)\\
        &=\left[\exp\left(i\epsilon\hat{K}_{\mathcal{G}} + o(\epsilon^2)\right)\right]^2\\
        &\cong \hat{V}_{\epsilon}^2= \left(\hat{U}_{\text{ideal}}\hat{U}_{\epsilon}^{\dagger}\right)^2 = \hat{U}_{\text{ideal}}\hat{U}_{\epsilon}^{\dagger}\cdot\hat{U}_{\text{ideal}}\hat{U}_{\epsilon}^{\dagger}
    \end{aligned}
\end{equation}
Left multiplying $\hat{U}_{\text{ideal}}^{\dagger}$ on both side, we obtain $\hat{U}_{2\epsilon}^{\dagger}\cong \hat{U}_{\epsilon}^{\dagger}\hat{U}_{\text{ideal}}\hat{U}_{\epsilon}^{\dagger}$, or equivalently,
\begin{equation}
    \hat{U}_{\text{ideal}} \cong \hat{U}_{\epsilon} \hat{U}_{2\epsilon}^{\dagger} \hat{U}_{\epsilon}.
\end{equation}
In the particular case of dealing with asymmetric error in trapped-ion systems, we have the explicit forms of $\hat{G}$ and $\hat{E}$, which are defined in the main text Sec.~\ref{subsec:dd} as
\begin{equation}
    \begin{aligned}
        \hat{G} &= \sigma_x^{j_1}\sigma_x^{j_2},\\
        \quad\hat{E} &= \frac{1}{2}\sum_{j=j_1,j_2}\sigma_z^{j}
    \end{aligned}
    \label{eq:GnE}
\end{equation}
satisfying $\{\hat{G}, \hat{E}\} = 0$, with which we can evaluate the derivative directly:
\begin{equation}
    \begin{aligned}
        &-i\frac{\partial}{\partial\epsilon}e^{-i\Theta\hat{G}+i\epsilon\hat{E}}|_{\epsilon=0}\\
        =& -i\frac{\partial}{\partial\epsilon}\left[ \mathbb{I} + (-i\Theta\hat{G}+i\epsilon\hat{E}) + \frac{1}{2!}(-i\Theta\hat{G}+i\epsilon\hat{E})^2 + \cdots\right]\bigg|_{\epsilon=0}\\
        =&\hat{E} + \frac{1}{3!}(-i\Theta\hat{G})^2\hat{E} + \frac{1}{5!}(-i\Theta\hat{G})^4\hat{E} + \cdots\\
        =&\frac{\hat{G}^{-1}}{2i\Theta} (e^{i\Theta\hat{G}}-e^{-i\Theta\hat{G}})\cdot\hat{E},
    \end{aligned}
\end{equation}
in the penultimate step we have applied the anti-commuting condition of $\hat{G}$ and $\hat{E}$. A closed form of $\hat{K}_{\mathcal{G}}$ is further given as
\begin{equation}
    \begin{aligned}
        \hat{K}_{\mathcal{G}} & = \hat{U}_{\text{ideal}}\cdot(-i\frac{\partial}{\partial\epsilon}\hat{U}^{\dagger}_{\epsilon}\big|_{\epsilon = 0})\\
        & = \frac{\hat{G}^{-1}}{2i\Theta} (e^{2i\Theta\hat{G}}-\mathbb{I})\cdot\hat{E}
    \end{aligned}
\end{equation}
By plugging in the definition of $\hat{G},\hat{E}$ (Eq. \ref{eq:GnE}) and $\Theta=\pi/4$, we secure the closed form of $\hat{K}_{\mathcal{G}}$ as
\begin{equation}
    \hat{K}_{\mathcal{G}} = \frac{2}{\pi}(\mathbb{I}+i\sigma_x^{j_1}\sigma_x^{j_2})(\sigma_z^{j_1}+\sigma_z^{j_2})
\end{equation}
It is easy to check that $e^{i\epsilon\hat{K}_{\mathcal{G}}}\hat{U}_{\epsilon} = \hat{U}_{\rm ideal} + o(\epsilon^2)$ by expanding all terms to leading order in $\epsilon$. 
\subsection{GBC scheme generalized to compensate higher order inaccuracies}
In the end of this part, we point out that although our method focuses mainly on eliminating the leading order error in $\epsilon$, it could be easily generalized to deal with higher order inaccuracies when necessitated by the scenario. In the previous derivation we have shown
$\hat{U}_{\epsilon}^{(gbc)}=\hat{U}_{\text{ideal}}+o(\epsilon^2)$, which could be regarded as the new ``noisy'' operator with leading order error in $o(\epsilon^2)$. Hence by similar arguments we suggest that
\begin{equation}
    \hat{U}_{\epsilon}^{(gbc)}\left(\hat{U}_{2\epsilon}^{(gbc)}\right)^{\dagger}\hat{U}_{\epsilon}^{(gbc)} = \hat{U}_{\epsilon} \hat{U}_{2\epsilon}^{\dagger} \hat{U}_{\epsilon} \cdot \hat{U}_{2\epsilon}^{\dagger} \hat{U}_{4\epsilon}\hat{U}_{2\epsilon}^{\dagger} \cdot \hat{U}_{\epsilon} \hat{U}_{2\epsilon}^{\dagger} \hat{U}_{\epsilon} = \hat{U}_{\text{ideal}} + o(\epsilon^4).
\end{equation}
If we define $\hat{U}^{(1)}_{\epsilon}\equiv\hat{U}_{\epsilon}^{(gbc)}$, and the recursive relation
\begin{equation}
    \hat{U}^{(k+1)}_{\epsilon}\equiv\hat{U}^{(k)}_{\epsilon}\left(\hat{U}^{(k)}_{2\epsilon}\right)^{\dagger}\hat{U}^{(k)}_{\epsilon},
\end{equation}
it is obvious that $\hat{U}^{(k)}_{\epsilon}$ serves as a k-th order approximation of $\hat{U}_{\text{ideal}}$ with error term $o(\epsilon^{2^k})$. This observation suggests that we can suppress the asymmetric error to an arbitrary level just by recursively applying our compensation scheme. However, the drawback of requiring extended total gatetime prevents us from doing so to an infinite order, since incoherent error would accumulate during gate operations and corrupt our state irreversibly. This detrimental effect could also be seen from the numerical results in Section.~\ref{sec:methods}, indicating a trade-off between suppressing the asymmetric error and limiting the gatetime.

\section{Amplitude Modulation for Robust 2-Qubit Gate}
\label{app:amplitude}
Starting with Eq.~\eqref{eq:2rd_mag},

\begin{equation}
\begin{aligned}
U(t)
=& \exp \Bigg( i \Theta(t) \sigma_x^{j_1} \sigma_x^{j_2} -\frac{i}{2} t \Delta\omega \sum_j \sigma_z^j \\
& -\frac{i}{2} \sum_{m j}\left(\alpha_m(t) a_m^{\dagger}+\alpha_m^*(t) a_m\right) \eta_{mj} \sigma_x^j \\
& -\frac{i}{4} \Delta\omega \sum_{m j}\left(\beta_m(t) a_m^{\dagger}+\beta_m^*(t) a_m\right) \eta_{mj} \sigma_y^j \Bigg),
\end{aligned}
\end{equation}
where
\begin{equation}
\begin{aligned}
\alpha_m(t) &= \int_0^t f(\tau) e^{i \delta_m \tau} \, d\tau, \\
\beta_m(t) &= \int_0^t a_m(\tau) \, d\tau + i \frac{\partial \alpha_m(t)}{\partial \delta_m}, \\
\Theta(t) &= \frac{1}{4}\sum_m \eta_{mj_1} \eta_{mj_2} \\
&\quad \int_0^t d\tau_1 \int_0^{\tau_1} 2f(\tau_1) f(\tau_2) \sin{(\delta_m(\tau_1-\tau_2))}.
\end{aligned}
\end{equation}

Here we use $f(\tau)$ to denote the amplitude of laser at time $\tau$. Amplitude modulation with robust conditions uses a waveform satisfying $f(t)=f(\tau-t)$ and $\overline{\alpha_m}=0$ for all $m$. As proven in Ref. \cite{leung2018robust}, this method decouples the phonons from spins at gate time $\tau$ ($\alpha_m(\tau)=0$) and makes the system insensitive to symmetric fluctuations in $\delta_m$ ($\frac{\partial \alpha_m(\tau)}{\partial \delta_m}=0$) while setting $\Theta=\Xi$. In our case, it eliminates $\beta$ and $\alpha$ at gate time $\tau$, while setting $\Theta=\Xi$:
\begin{equation}
\alpha_m(\tau)=0,\quad \beta_m(\tau)=0,\quad \Theta(\tau)=\Xi.
\end{equation}

To understand this result, we express $\beta_m(\tau)$ in terms of $\alpha_m$:
\begin{equation}
\begin{aligned}
\beta_m(\tau) & =\int_0^\tau d t \alpha_m(t) d t - \int_0^\tau f(t) t e^{i \tilde{\delta}_m t} d t \\
& = \tau \space \overline{\alpha_m} - \tau \alpha_m(\tau) + \int_0^\tau \alpha_m(t) d t \\
& = 2 \tau \space \overline{\alpha_m} - \tau \alpha_m(\tau).
\end{aligned}
\end{equation}
Applying the pulse symmetry constraint $f(t)=f(\tau-t)$, we obtain:
\begin{equation}
\alpha_m(\tau-t)=\alpha_m(\tau) - e^{i \tilde{\delta}_m t} \alpha_m^*(t).
\end{equation}
Integrating this equation over the interval $[0, \tau]$ yields $\alpha_m(\tau)=e^{i \tilde{\delta}_m \tau} \overline{\alpha_m^*}$. This implies that under the pulse symmetry constraint, setting $\overline{\alpha_m}=0$ also results in $\alpha_m(\tau)=0$. Thus, when we impose both conditions $f(t)=f(\tau-t)$ and $\overline{\alpha_m}=0$, we simultaneously satisfy the constraints on $\alpha_m(\tau)$ and $\beta_m(\tau)$. 

Define $f(t)$ as a linear composition of piecewise functions, 
\begin{equation}
\begin{aligned}
f(t) &= \sum_{i=0}^{i<N_{\text{seg}}} x_i f_i(t) \\
f_i(t) &= \begin{cases}
1, & \text{if } \frac{i\tau}{N_{\text{seg}}} < t < \frac{(i+1)\tau}{N_{\text{seg}}} \\
0, & \text{otherwise}
\end{cases}
\end{aligned}
\label{eq:piecewise}
\end{equation}

With Eq.~\eqref{eq:piecewise}, the waveform can be represented by an vector $\mathbf{x}$, and the constraints can be expressed as linear equations of $\mathbf{x}$:
\begin{equation}
A\mathbf{x}=\mathbf{0}
\end{equation}
\begin{equation}
\mathbf{x}^T M \mathbf{x}=\Xi
\end{equation}
where
\begin{equation}
A_{ml}=\int_0^\tau dt_1 \int_0^{t_1} f_l(t_2) e^{i \delta_m t_2} d t_2
\end{equation}
\begin{equation}
M_{ij}=-\frac{1}{4}\sum_{m}\eta_{m,j_1}\eta_{m,j_2}\int^\tau_0 dt_1 \int_0^{t_1} dt_2 (f_i(t_1) f_j(t_2) + f_i(t_2) f_j(t_1)) \sin(\delta_m (t_2 - t_1))
\end{equation}

An $\mathbf{x}$ satisfying the constraints should be written as $\mathbf{x}=\ker(A)\mathbf{v}$. By finding a $\mathbf{v}$ which satisfies $\mathbf{v}^T \ker(A)^T M \ker(A) \mathbf{v}=\Xi$, we get an $\mathbf{x}$ that meets the two constraints.

To generate the waveform with the minimum intensity, we diagonalize the matrix $M$, subsequently extracting the eigenvalue $\lambda$ with the largest absolute value that shares the same sign of positivity or negativity as $\Xi$, along with its corresponding eigenvector $\mathbf{v}_{\lambda}$. Thus, we have:
\begin{equation}
\mathbf{x}=\ker(A)\mathbf{v}=\sqrt{\frac{\Xi}{\lambda}}\ker(A)\mathbf{v}_{\lambda}
\end{equation}

\section{Introduction of Asymmetrix Errors in Stimulated Raman Transition Gates}
\label{app:SR_error}
Consider an $N$-ion chain confined in a Paul trap. Each ion is treated as a three-level system with $\ket{0}$ and $\ket{1}$ defining the hyperfine "clock" states used for qubit states, and $\ket{e}$ as the excited state used in stimulated Raman transitions. The Hamiltonian of the system is given by
\begin{equation}
    H = \sum_{i=1}^{N}(\omega_{01}\ket{1}_i\bra{1}+\omega_{0e}\ket{e}_i\bra{e})+\sum_k \omega_k a_k^\dagger a_k,
\end{equation}
where $\omega_{01}$ and $\omega_{0e}$ are the energy differences between the qubit states and the excited state, respectively, and $\omega_k$ is the frequency of the $k$-th mode of the harmonic oscillator. The system interacts with a bichromatic laser field applied to ion $j$ with frequencies $\omega_{1}$ and $\omega_{2}$, and Rabi frequencies $\Omega_{1}$ and $\Omega_{2}$, respectively. The light couples $\ket{0}$ and $\ket{e}$, and $\ket{1}$ and $\ket{e}$ off-resonantly. The light field Hamiltonian is given by
\begin{equation}
    H^\prime = \Omega_1 \cos(\mathbf{k}_1 \cdot \mathbf{r}_j - \omega_1 t - \phi_1)(\ket{0}_j\bra{e} + \ket{e}_j\bra{0})
    + \Omega_2 \cos(\mathbf{k}_2 \cdot \mathbf{r}_j - \omega_2 t - \phi_2)(\ket{1}_j\bra{e} + \ket{e}_j\bra{1}),
\end{equation}
where $\mathbf{k}_1$ and $\mathbf{k}_2$ are the wave vectors of the two lasers, $\mathbf{r}_j$ is the position of ion $j$, and $\phi_1$ and $\phi_2$ are the phases of the two lasers. Under the interaction picture defined by 
\begin{equation}
    H_0 = \sum_{i \neq j}(\omega_{01}\ket{1}_i\bra{1} + \omega_{0e}\ket{e}_i\bra{e}) + \sum_k \omega_k a_k^\dagger a_k
    + \omega_{01}\ket{1}_j\bra{1} + \omega_{1}\ket{e}_i\bra{e},
\end{equation}
and under the rotating-wave approximation, the interaction Hamiltonian becomes
\begin{equation}
    H_{\text{int}} = -\Delta\ket{e}_j\bra{e} + \frac{\Omega_1}{2}(\exp(i(\mathbf{k}_1 \cdot \mathbf{r}_j - \phi_1))\ket{e}_j\bra{0} + h.c.) + \frac{\Omega_2}{2}(\exp(i(\mathbf{k}_2 \cdot \mathbf{r}_j + \delta t - \phi_2))\ket{e}_j\bra{1} + h.c.),
\end{equation}
where we define $\Delta = \omega_{1} - \omega_{0e}$ and $\delta = \omega_1 - \omega_2 - \omega_{01}$. In the case of $\left|\Delta\right| \gg \left|\Omega_1\right|, \left|\Omega_2\right|, \left|\delta\right|$, the excited state can be adiabatically eliminated, and the effective Hamiltonian is given by 
\begin{equation}
    H_{\text{int}} = \frac{\Omega_1 \Omega_2}{4 \Delta} \exp(i(-\Delta \mathbf{k} \cdot \mathbf{r}_j + \delta t + \Delta \phi))\ket{0}_j\bra{1} + h.c.
\end{equation}
Under the phase-insensitive geometry, where three lasers are applied to the ion, the effective Hamiltonian of the system is the summation of two stimulated Raman transitions driven by $\Omega_1$, $\Omega_2$ and $\Omega_3$, $\Omega_1$, respectively. The first laser has a frequency $\omega_1$ and direction $\mathbf{k}_1$. The second and third lasers have the same direction $\mathbf{k}_2$, with frequencies red-detuned and blue-detuned relative to $\omega_1$, respectively, such that $\omega_2 = \omega_1 - \omega_{01} - \mu$ and $\omega_3 = \omega_1 + \omega_{01} - \mu$.
\begin{align}
    H_{\text{int}} = &\left(\frac{\Omega_1 \Omega_2}{4 (\omega_1-\omega_{0e})} \exp(i(-\Delta \mathbf{k} \cdot \mathbf{r}_j + \mu t + \Delta \phi_{12})) \right. 
    \quad + \left. \frac{\Omega_1 \Omega_3}{4 (\omega_3-\omega_{0e})} \exp(i(\Delta \mathbf{k} \cdot \mathbf{r}_j - \mu t - \Delta \phi_{13}))\right)\ket{0}_j\bra{1} + \text{h.c.} \notag \\
    &+ E_0\ket{0}_j\bra{0} + E_1\ket{1}_j\bra{1}, \label{eq:sr_ham_intermidate}
\end{align}
where $\Delta \phi_{12} = \phi_1 - \phi_2$ and $\Delta \phi_{13} = \phi_1 - \phi_3$. $E_0 = \frac{\Omega_1^2}{4(\omega_1 - \omega_{0e})} + \frac{\Omega_3^2}{4(\omega_3 - \omega_{0e})}$, $\Delta \mathbf{k} = \mathbf{k}_1 - \mathbf{k}_2$ and $E_1 = \frac{\Omega_2^2}{4(\omega_1 - \omega_{0e})} + \frac{\Omega_1^2}{4(\omega_3 - \omega_{0e})}$. The phase-insensitive geometry requires that 
\begin{equation}
    \Omega_j = \frac{\Omega_1 \Omega_2}{4  (\omega_1-\omega_{0e})} = \frac{\Omega_1 \Omega_3}{4 (\omega_3-\omega_{0e})},
\end{equation}
thus we have $E_0 = \frac{\omega_3 - \omega_{0e}}{\omega_1 - \omega_{0e}} E_1$. The Hamiltonian can be further simplified as
\begin{equation}
    H_{\text{int}} = \Omega_j \cos(-\Delta \mathbf{k} \cdot \mathbf{r}_j + \mu t + \phi_j^{(m)})\left(e^{i \phi_j^{(s)}} \sigma_-^j + e^{-i \phi_j^{(s)}} \sigma_+^j\right) + \frac{\Delta E}{2} \sigma_z^j,
    \label{eq:phase-insensitive-ham}
\end{equation}
Here, $\phi_j^{(m)} = \phi_1 - \frac{\phi_{2} + \phi_{3}}{2}$ and $\phi_j^{(s)} = \frac{\phi_{3} - \phi_{2}}{2}$, with $\sigma_-^j = \ket{0}_j\bra{1}$ and $\sigma_+^j = \ket{1}_j\bra{0}$. Since lasers 2 and 3 are generated by modulating the same beam, their fluctuations in optical path difference are minimal. Therefore, we can approximate $\phi_j^{(s)} \approx 0$, which is the underlying principle of phase-insensitive geometry.

By redefining the zero point of the potential energy, the differential Stark shift is given by $\Delta E = E_1 - E_0$. This indicates that a differential Stark shift inherently exists in the phase-insensitive geometry. 

In principle, the differential Stark shift can be compensated by adjusting the laser frequencies. However, in amplitude-modulated gates, each segment corresponds to a different light intensity, leading to distinct energy shifts. Consequently, individual adjustments become costly. Moreover, other experimental factors, such as stray magnetic fields, may induce drifts in the qubit energy levels, which also produce asymmetric errors. 

Apart from internel frequency drifts, asymmetric errors can also result from laser miscalibration. Unlike laser miscalibrations in Sec.~\ref{sec:MS_gate}, where both laser frequencies shift equally, in phase-insensitive geometry, asymmetric error arises due to the opposite shifts in $\omega_2$ and $\omega_3$, which may stem from errors in the modulation process. Specifically, the laser frequencies are related as:
\begin{equation}
    \omega_3 = \omega_1 + \omega_{01} - \mu + \lambda_l, \quad \omega_2 = \omega_1 - \omega_{01} - \mu - \lambda_l.
\end{equation}
Under interaction picture defined by
\begin{equation}
    H_0^\prime = \sum_{i \neq j} \left[ (\omega_{01} + \lambda_l) \ket{1}_i\bra{1} + \omega_{0e} \ket{e}_i\bra{e} \right] + \sum_k \omega_k a_k^\dagger a_k
    + \omega_{01} \ket{1}_j\bra{1} + \omega_1 \ket{e}_i\bra{e},
\end{equation}
the interaction Hamiltonian becomes:
\begin{equation}
    H_{\text{int}} = \Omega_j \cos(-\Delta \mathbf{k} \cdot \mathbf{r}_j + \mu t + \phi_j^{(m)})\sigma_n^j + \lambda \sigma_z^j,
    \label{eq:exact_ham_sr}
\end{equation}
where we define $\sigma_n^j= e^{i \phi_j^{(s)}} \sigma_-^j + e^{-i \phi_j^{(s)}} \sigma_+^j $. $\lambda = \lambda_l + \frac{\Delta E}{2}$ represents the overall asymmetric error introduced by internal energy drifts and laser frequency miscalibration. Note that $\lambda_l$ is incorporated into $H_0^\prime$, thus in this frame, the state $\ket{1}$ undergoes a rotation by $e^{i\lambda_l t}$ relative to the frame defined by $H_0$. This results in the accumulation of a relative phase. Phase modulation is required in subsequent gate operations to maintain consistency with this frame transformation.

The quantization of ion position gives
\begin{equation}
    \Delta \mathbf{k} \cdot \mathbf{r}_j = \sum_m \eta_m b_j^m(a_m e^{-i\omega_m t} + a_m^\dagger e^{i\omega_m t})
\end{equation}
where $\eta_m = \Delta \mathbf{k}_z \sqrt{\frac{\hbar}{2M\omega_m}}$ is the Lamb-Dicke parameter. Under the Lamb-Dicke regime, $\eta_m$ is small, thus we can expand Eq.~\eqref{eq:exact_ham_sr} in powers of $\eta_m$:
\begin{equation}
H_{\text{int}}=\Omega_j cos(\mu t+ \phi_j^{(m)})   \sigma^j_{n} + \Omega_j sin(\mu t+ \phi_j^{(m)}) \sum_m \eta_m b_j^m(a_m e^{-i\omega_m t} + a_m^\dagger e^{i\omega_m t})\sigma^j_{n}  +\lambda \sigma_z^j+O(\eta_m^2)
\end{equation}
Under the condition $|\omega_m - \mu| \ll \mu, \omega_m+\mu$, rotating wave approximation can be applied, which simplifies the Hamiltonian to:
\begin{equation}
    H_{\text{int}}\approx \frac{\Omega_j}{2i}\sum_m\left(a_m e^{-i\delta_m t+\phi_j^{(m)}} -a_m^\dagger e^{i\delta_m t-\phi_j^{(m)}} \right) \sigma^j_{n} +\lambda \sigma_z^j
\end{equation}
To implemente the MS-gate, light is shined on two ions. By setting $\phi_j^{(m)}=\frac{\pi}{2}$, $ \phi_j^{(s)}=0$ for both ions, and summing over ions $j_1$ and $j_2$ with $\Omega_{j_1}=\Omega_{j_2}$, we recover the Hamiltonian with asymmetric errors Eq.~\eqref{eq:miscalib_ham} in Sec.~\ref{sec:MS_gate}.

\section{Supplementary Numerical Results}
\label{app:sup_num}
\begin{figure}[H]
    \centering
    \includegraphics[width=3.4in]{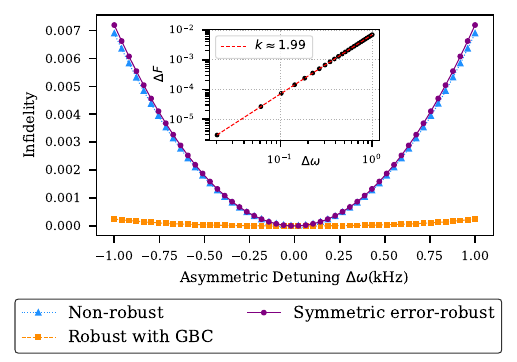}
    \caption{Performance comparison against varying asymmetric detuning \(\Delta \omega\) for three protocols evaluated via numerical simulation of a three-ion \(\mathrm{^{40}Ca^+}\) system. A target entangling angle of \(\frac{\pi}{4}\) is set for the first two ions from left to right. Gate infidelity as a function of asymmetric detuning \(\\Delta \omega\) is shown for all three methods: non-robust waveform, symmetric error-robust waveform, and robust waveform with GBC.}
    \label{fig:performance_comparison_n=3}
\end{figure}

Our gate scheme is directly applicable to multi-ion chains, but simulating their dynamics poses challenges due to the non-commutative terms in the Hamiltonian (Eq.~\eqref{eq:miscalib_ham}). In the ideal Hamiltonian (Eq.~\eqref{ideal_ham}), phonon terms commute pairwise, enabling efficient simulations with a Hilbert space dimensionality of \(4N_c\), where \(N_c\) is the cutoff for each phonon. However, asymmetric errors disrupt this commutation, increasing simulation complexity to \(O((N_c^{N_i})^3)\), where \(N_i\) is the number of ions. This limits us to \(N_i \leq 3\), but the results for \(N_i=3\) agree with theoretical predictions, as shown in Fig.~\ref{fig:performance_comparison_n=3}.

\end{document}